\documentclass{aa}

\usepackage{graphicx}
\usepackage{CJK}
\usepackage{enumitem}
\usepackage[varg]{txfonts}
\usepackage{amsmath}
\usepackage{amssymb}
\usepackage{footmisc}
\usepackage{booktabs}
\usepackage[colorlinks = true,
            linkcolor = blue,
            urlcolor  = blue,
            citecolor = blue,
            anchorcolor = blue,
            unicode]{hyperref}

\begin{document}

\begin{CJK*}{UTF8}{gbsn}

\title{Evolved Massive Stars at Low-metallicity \uppercase\expandafter{\romannumeral4}.\\ Using 1.6 $\mu$m ``H-bump'' to identify red supergiant stars:\\ a case study of NGC 6822}
\titlerunning{Evolved massive stars at low-Z \uppercase\expandafter{\romannumeral4}. Identifying RSGs in NGC 6822}

\author{
Ming Yang (杨明) \inst{1} \and Alceste Z. Bonanos \inst{1} \and Biwei Jiang (姜碧沩) \inst{2} \and Man I Lam (林敏仪) \inst{3} \and Jian Gao (高健) \inst{2} \and Panagiotis Gavras \inst{4} \and Grigoris Maravelias \inst{1,5} \and Shu Wang (王舒) \inst{6} \and Xiao-Dian Chen (陈孝钿) \inst{6} \and Frank Tramper \inst{1} \and Yi Ren (任逸) \inst{2} \and Zoi T. Spetsieri \inst{1}
}
\authorrunning{Yang, Bonanos \& Jiang et al.}

\institute{
IAASARS, National Observatory of Athens, Vas. Pavlou and I. Metaxa, Penteli 15236, Greece\\
                \email{myang@noa.gr} \and
Department of Astronomy, Beijing Normal University, Beijing 100875, People's Republic of China \and
Key Laboratory of Space Astronomy and Technology, National Astronomical Observatories, Chinese Academy of Sciences, Beijing 100101, People's Republic of China \and
Rhea Group for ESA/ESAC, Camino bajo del Castillo, s/n, Urbanizacion Villafranca del Castillo, Villanueva de la Cañada, 28692 Madrid, Spain \and
Institute of Astrophysics, Foundation for Research and Technology-Hellas, Heraklion 71110, Greece \and
CAS Key Laboratory of Optical Astronomy, National Astronomical Observatories, Chinese Academy of Sciences, Datun Road 20A, Beijing 100101, People's Republic of China 
}

\abstract{
We present a case study of using a novel method to identify red supergiant (RSG) candidates in NGC 6822, based on their 1.6 $\mu$m ``H-bump''. We collected 32 bands of photometric data for NGC 6822 ranging from optical to mid-infrared (MIR), derived from Gaia, PS1, LGGS, VHS, UKIRT, IRSF, HAWK-I, Spitzer, and WISE. By using the theoretical spectra from MARCS, we demonstrate that there is a prominent difference around 1.6 $\mu$m (``H-bump'') between low-surface-gravity (LSG) and high-surface-gravity (HSG) targets. Taking advantage of this feature, we identify efficient color-color diagrams (CCDs) of rzH (r-z versus z-H) and rzK (r-z versus z-K) to separate HSG (mostly foreground dwarfs) and LSG targets (mainly background red giant stars (RGs), asymptotic giant branch stars (AGBs), and RSGs) from crossmatching of optical and near-infrared (NIR) data. Moreover, synthetic photometry from ATLAS9 also give similar results. Further separating RSG candidates from the rest of the LSG candidates as determined by the ``H-bump'' method is done by using semi-empirical criteria on NIR color-magnitude diagrams (CMDs), where both the theoretic cuts and morphology of the RSG population are considered and resulted in 323 RSG candidates. Meanwhile, the simulation of foreground stars from Besan\c{c}on models also indicates that our selection criteria is largely free from the contamination of Galactic giants. In addition to the ``H-bump'' method, we also use the traditional BVR method (B-V versus V-R) as a comparison and/or supplement, by applying a slightly aggressive cut to select as much as possible RSG candidates (358 targets). Furthermore, the Gaia astrometric solution is used to constrain the sample, where 181 and 193 targets were selected from the ``H-bump'' and BVR method, respectively. The percentages of selected targets in both methods are similar as $\sim$60\%, indicating the comparable accuracy of the two methods. In total, there are 234 RSG candidates after combining targets from both methods with 140 ($\sim$60\%) of them in common. The final RSG candidates are in the expected locations on the MIR CMDs with $[3.6]-[4.5]\lesssim0$ and $J-[8.0]\approx1.0$. The spatial distribution is also coincident with the far-ultraviolet-selected star formation regions, suggesting the selection is reasonable and reliable. We indicate that our method also can be used to identify other LSG targets like RGs and AGBs, as well as applied to most of the nearby galaxies by utilizing the recent large-scale ground-based surveys. Future ground and space facilities may promote its application beyond the Local Group.
}


\maketitle

\section{Introduction}

Red supergiant stars (RSGs) are helium-fusing, evolved massive stars with initial mass about 7$\sim$25 $M_{sun}$ \citep{Levesque2005, Ekstrom2013}. They represent an extremity of stellar evolution as the coldest and largest members of the massive star population. Their mass-loss rate (MLR) determines their lifetime and how they end as hydrogen-rich Type \uppercase\expandafter{\romannumeral2}-P supernovae (SN), or exit the RSG phase during core He and shell H fusion and evolve backwards to the blue end of the Hertzsprung-Russell (H-R) diagram producing a ``blue loop'' \citep{Massey2013, Smith2014}. Notice that, currently, the underlying physics of the ``blue loop'' is still largely uncertain. One explanation involves the opposite ``mirror effect'' with expanding He-core and contracting H-envelope, but (as with other evolutionary stages) it may also be related to other factors as treatments of rotation, mixing, overshooting, metallicity, and MLR \citep{Meynet2015}. RSGs are also expected to be in a key stage for the formation of interacting binaries \citep{Neugent2020}. With their high luminosities ($\gtrsim10^4 L_\sun$) and consequently bright magnitudes in the optical and near-infrared (NIR) bands, they can be easily observed beyond our Milky Way (MW; \citealt{Bergemann2012}). With such an important role in the stellar physics, the investigation of physical properties and evolution of RSGs has achieved large progress in the past half century \citep{Humphreys1979, Massey2003, Gonzalez2015, Davies2017}. 

To better understand the nature of the RSGs, it is crucial to build a representative sample covering wide ranges of both metallicity and luminosity. However, the major challenge of identifying extragalactic RSGs is the foreground contamination from galactic dwarfs and giants. \citet{Massey1998} adopted a simple but effective way to distinguish between foreground high-surface-gravity (HSG) dwarfs and background low-surface-gravity (LSG) supergiants, by using color-color diagram (CCD) of V-R versus B-V (BVR) in the Johnson filter set. It was mainly based on the line-blanketing effect, which was particularly notable in the B-band, due to a number of weak metal lines in the regime. For stars with LSG (e.g., RSGs), weak metal absorption lines decrease flux in the B-band proportionally more than for stars with HSG (e.g., dwarfs), and so it is possible to differentiate LSG and HSG stars by comparing relative flux in the B-band. However, as future observations are gradually moving away from the traditional Johnson filter set (and exploring non-optical wavelength regimes), the modern filter sets have different effective wavelengths and coverages other than the Johnson system. In that sense, similar color-color diagrams (CCDs) do not resemble the same effect of the BVR diagram. Thus, it is necessary to find new combinations of colors to separate LSGs from HSGs, in order to utilize the data from the most recent large-scale surveys.

We selected NGC 6822 for a case study due to its rich reservoir of multiwavelength data. As a member of the Local Group (LG), NGC 6822 is an isolated, dwarf irregular galaxy (dIrr), which is similar to the Small Magellanic Cloud (SMC) in both size and metallicity \citep{Tolstoy2001, Cioni2005, McConnachie2012, GarciaRojas2016}. With a distance modulus about $23.40\pm0.05$ \citep{Feast2012, Rich2014}, it is one of the nearest dwarf galaxy in the LG. Moreover, previous studies (e.g., \citealt{Levesque2012, Patrick2015}) have spectroscopically identified several RSGs in NGC 6822, making it a perfect testbed for our purpose. In this paper, we present our new method based on the ``H-bump'' around 1.6 $\mu$m, which is a distinct humped spectral feature of LSG targets caused by an apparently smaller opacity in the H$^-$ of LSGs than HSGs, to identify RSG candidates in NGC 6822. The multiwavelength data and identification of RSG candidates are presented in \textsection2 and \textsection3, respectively. The evaluation of the candidates is described in \textsection4. The summary is given in \textsection5.

\section{Multiwavelength data for NGC 6822}

We collected multiwavelength data in 32 bands ranging from optical to the mid-infrared (MIR) for NGC 6822. The MIR data were collected from \citet{Khan2015} where Spitzer/IRAC [3.6], [4.5], [5.8], [8.0], and MIPS [24] point-source catalogs for seven galaxies were presented, and ALLWISE catalog \citep{Cutri2013} from Wide-field Infrared Survey Explorer (WISE; \citealt{Wright2010}). The NIR data were collected from \citet{Sibbons2012} with deep, high quality JHK photometry taken by WFCAM on the United Kingdom Infra-Red Telescope (UKIRT), Visible and Infrared Survey Telescope for Astronomy (VISTA) Hemisphere Survey (VHS) Data Release 6 (DR6) \citep{McMahon2013}, \citet{Whitelock2013}. with 3.5-yr survey of the central regions of the NGC 6822 taken by the Infrared Survey Facility (IRSF) equipped with the SIRIUS camera, and \citet{Libralato2014} with high-precision $JK_{\rm S}$ photometry taken by High Acuity Wide-field K-band Imager (HAWK-I) on the ESO Very Large Telescope (VLT). The optical photometric data were collected from two large surveys of Panoramic Survey Telescope and Rapid Response System (Pan-STARRS, PS1; \citealt{Chambers2016}) DR2 and Gaia DR2 \citep{Gaia2016, Gaia2018}, and one dedicated survey of Local Group Galaxy Survey (LGGS; \citealt{Massey2007}) for observing nearby galaxies currently forming stars. 

\begin{figure}
\center
\includegraphics[scale=0.5]{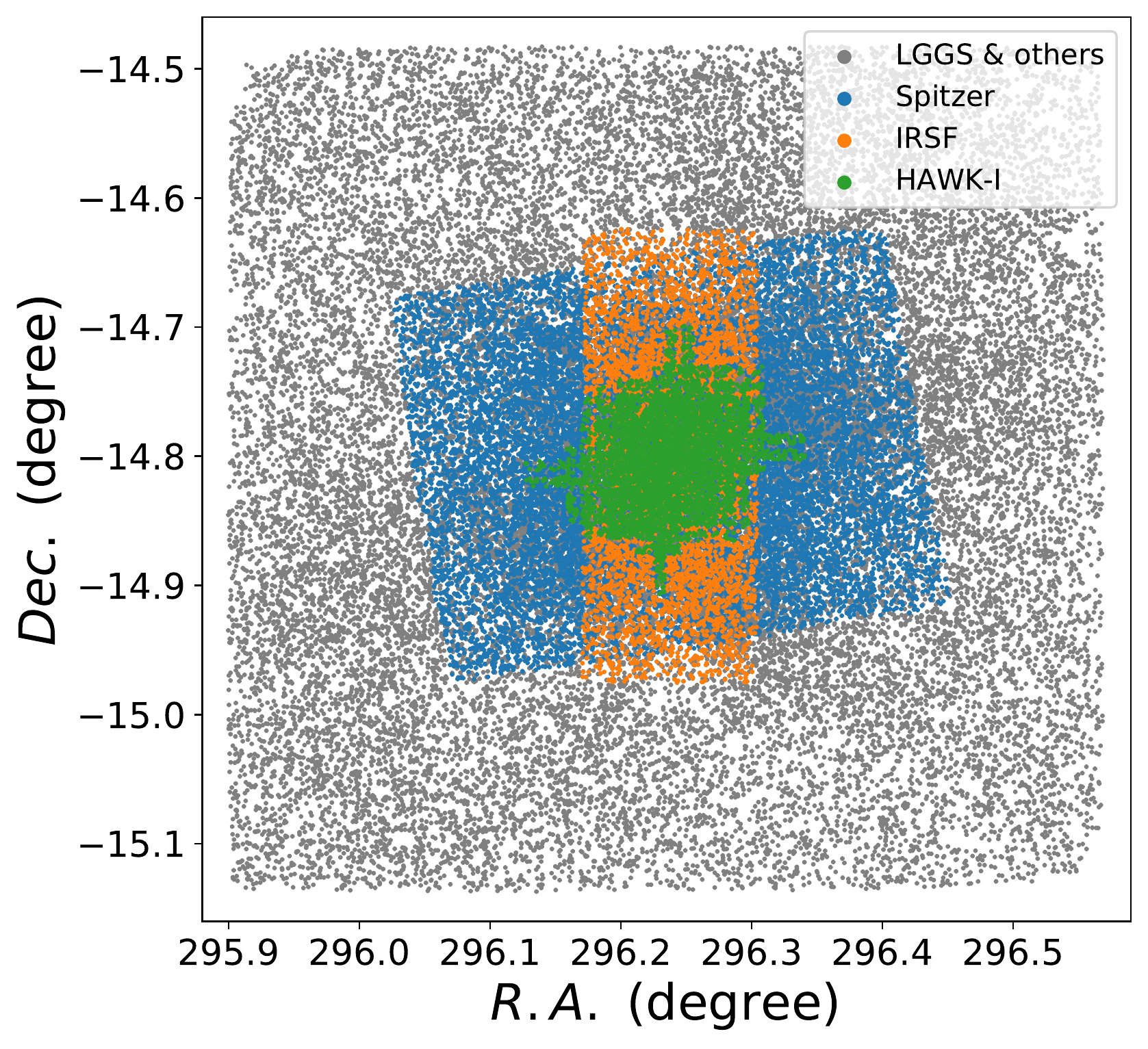}
\caption{The adopted common observational region of NGC 6822 based on the LGGS. Different colors indicate different surveys.
\label{ngc6822_spatial}}
\end{figure}

As different surveys covering different field of views (FoVs), we tended to define a common region overlapped by all the surveys as much as possible, without excessive sky coverage. Limiting the FoV would allow us to largely reduce the total foreground contamination. Figure~\ref{ngc6822_spatial} shows the common region based on the LGGS, that all other surveys either share the same area or are located within it. Compared to the original dataset of \citet{Sibbons2012}, it was only $\sim$13.5\% of the sky area and $\sim$17.3\% of total number of targets, for which the overwhelming majority of foreground stars were excluded.

In order to reduce the false positive rate in the following analysis, following the procedure of \citet{Yang2019}, each collected dataset was preprocessed and self-deblended with a search radius of $2''$ as shown below (in each dataset, targets with neighbors within $2''$ were excluded). The deblending radius of $2''$ was justified based on the moderate angular resolution of $\sim$$1.8''$ of Spitzer. A search radius of $1''$ was used for further crossmatching between different datasets. 
\begin{itemize}[noitemsep,topsep=0pt,parsep=0pt,partopsep=0pt] 
\item 28,824 targets from \citet{Khan2015}, where all of them having 3-$\sigma$ detection in [3.6] and [4.5] bands, but not necessarily in [5.8], [8.0], and [24] bands.
\item 7,464 targets from ALLWISE catalog in the common region, with signal to noise ratio (S/N) larger than 3 in both [3.4] and [4.5] bands, but not necessarily in [12] and [22] bands. 
\item 25,848 targets from \citet{Sibbons2012} in the common region, with flags of -1 (stellar) or -2 (probably stellar) and photometric errors $\leq$0.3 mag in JHK bands. 
\item 16,903 targets from VHS DR6 with -3$~\leq~$mergedClassStat$~\leq~$3 (merged stellarness-of-profile statistic), mergedClass$~=-1$ (stellar) or -2 (probably stellar), and photometric errors $\leq$0.3 mag in $JK_{\rm S}$ bands. 
\item 5,657 targets from \citet{Whitelock2013} (only deblended with no other constraints).
\item 4,318 targets from \citet{Libralato2014}, with flags of $Jw(Kw)\neq0$ (weed-out flag with 0 being rejected), $q\_Jmag(q\_Kmag)\leq0.25$ (quality of point-spread-function (PSF) fit), and photometric errors $\leq$0.3 mag in $JK_{\rm S}$ bands. 
\item 23,977 targets from PS1 DR2, with flags of $nDetections>2$ (number of single epoch detections in all filters), $qualityflag<64$ (flag denoting whether this object is real or a likely false positive), $g(r, i, z)QfPerfect>0.75$ (maximum PSF weighted fraction of pixels totally unmasked from g(r, i, z) filter detections), and $rmeanpsfmag-rmeankronmag<0.05$ (a rough cut to separate stars and galaxies).
\item 22,826 targets from Gaia DR2 (only deblended with no other constraints).
\item 43,584 targets from LGGS, with photometric errors $\leq$0.3 mag in B, V, and R bands.
\end{itemize}

The catalogs for each dataset are listed in the Appendix~\ref{appendix1}. Notice that, there is a general problem for joining multiple catalogs that, a large number of false positive matches may appear between targets listed in different surveys. Thus, we did not intend to join all catalogs into a common, master catalog, but leave the decision to readers. This is actually a complicated technical issue and a natural outcome of our (or any other) multiwavelength catalogs. To put it simply, it is mainly due to three reasons that we prefer not to build a master catalog. The first reason is the lack of reference catalogs in NGC 6822. Unlike \citet{Yang2019, Yang2020b} that we built up source catalogs based on Gaia and Spitzer to study bona-fide dusty massive stars in the Magellanic Clouds, there is no comprehensive and reliable reference catalog for NGC 6822. For example, Gaia data is insufficient to determine the membership of a considerable part of targets in NGC 6822 as mentioned in \textsection4, while Spitzer data is dominated by the background galaxies at the faint magnitude (e.g. [3.6]$>$16 mag; \citealt{Ashby2009, Williams2015}). Secondly, different surveys are not necessarily observing the same population of targets, due to different wavelengths, sensitivities, depths, sky coverages, quality cuts, and so on. For example, dusty targets may be extremely faint and undetected at short wavelengths (e.g., B- or even V-band), but are very bright at long wavelengths (e.g., IR bands). Even for the similar wavelengths, some targets may be available in one survey but not necessarily in others due to, for example, different exposure times and quality cuts. Thirdly, indeed, we could build up a master catalog (we have actually tried it) by simply joining all catalogs by using their coordinates, for which each match represents a group of ``friends of friends''. However, it is important to notice that, for any particular pair in a matched group, there is no guarantee that the two objects match each other, but only that it can hop from one to the other via pairs which do match\footnote{http://www.star.bris.ac.uk/$\sim$mbt/stilts/sun256/matchGroup.html}. In that sense, it actually creates false matches, especially at faint magnitudes and crowded regions. Meanwhile, as our goal is to study RSGs, false positive is generally not a big issue, since RSGs are bright and we have deblended each dataset.

\section{Identifying red supergiant stars by using ``H-bump'' at 1.6 $\mu$m}

\begin{figure}
\center
\includegraphics[scale=0.3]{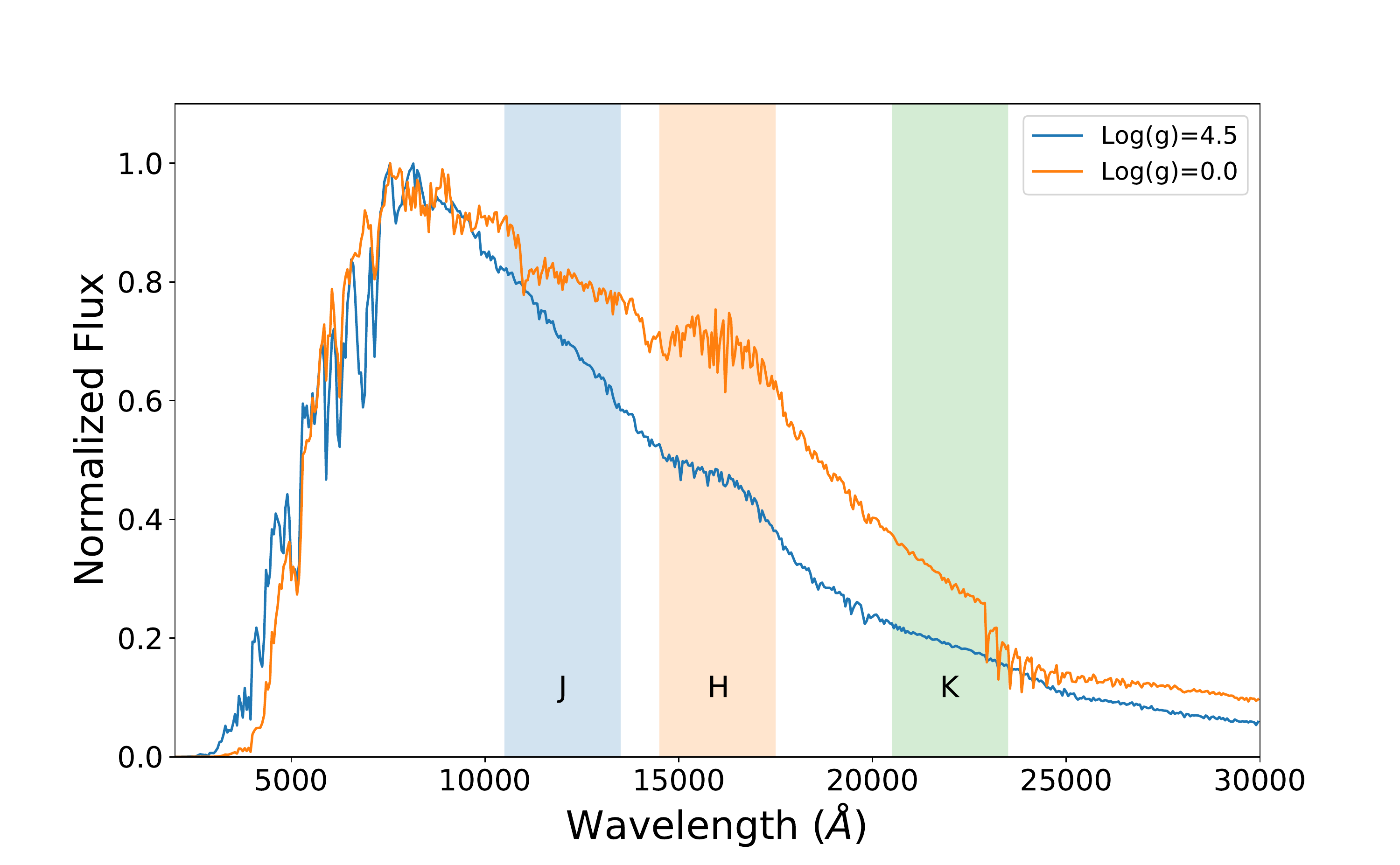}
\caption{Two normalized resampled MARCS spectra with the same $T_{\rm eff}=3700$ K at metallicity of -0.75, but for different surface gravities of 4.5 (HSG) and 0.0 (LSG). LSG target shows a humped spectral feature around 1.6 $\mu$m (``H-bump''). The approximate wavelength ranges of $JHK_{\rm (S)}$ filters are shown as different shades.
\label{ngc6822_logg}}
\end{figure}

To identify RSGs in NGC 6822, firstly, we used the theoretical spectra from one-dimensional, hydrostatic, spherical LTE model atmospheres of MARCS \citep{Plez2003, Gustafsson2003, Gustafsson2008} to demonstrate the general separation between HSG (e.g., dwarfs) and LSG (e.g, supergiants) targets. MARCS models were widely used in recent studies to fit moderate-resolution spectrophotometry of RSGs in the MW, the MCs, M31 and so on \citep{Levesque2005, Levesque2006, Massey2009, Davies2013}. Figure~\ref{ngc6822_logg} shows two normalized resampled spectra with the same effective temperature ($T_{\rm eff}$) of 3700K at metallicity (Z) of -0.75 but for different surface gravities ($Log~g$) of 4.5 (HSG) and 0.0 (LSG), respectively. From the diagram, it can be seen that there is an obvious difference ($\sim$25\% of normalized flux) between the HSG and LSG targets due to a humped spectral feature of LSG targets around 1.6 $\mu$m (``H-bump''). This ``H-bump'' is expected to appear in cool stars because of an apparently smaller opacity in the H$^-$ of supergiants than dwarfs (though they are both in a minimum of H$^-$ opacity). However, it will be suppressed at high metallicities by molecular absorption (e.g., CO; \citealt{John1988, Davies2013}). Thus, in principal, a color combination involving at least one of the $JHK_{(\rm S)}$ filters (as shown in the diagram) can be used to distinguish between foreground dwarfs and background supergiants. Notice that, the dwarfs in NGC6822 have no influence on the result, since they are much fainter than foreground dwarfs and supergiants in the same galaxy.

\begin{figure*}
\center
\includegraphics[scale=0.47]{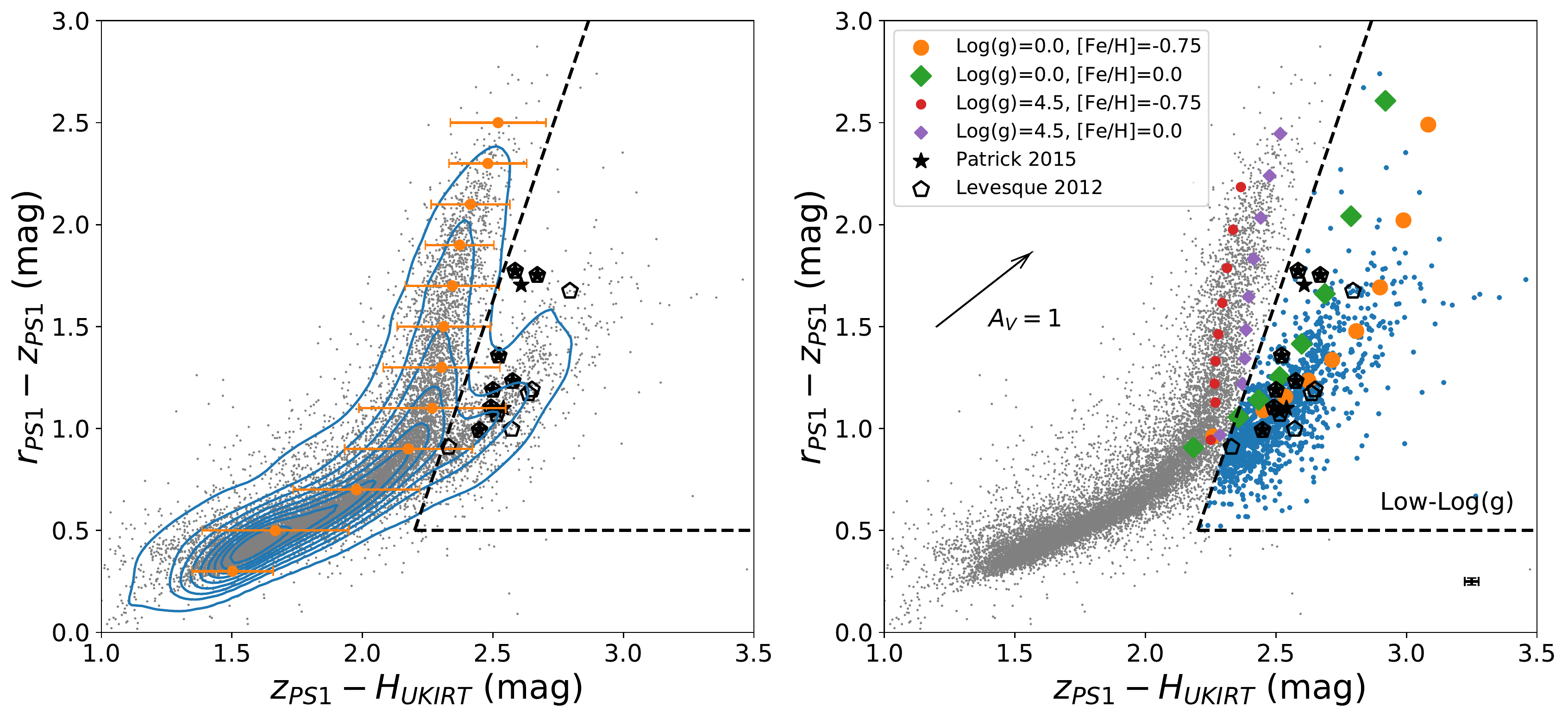}
\includegraphics[scale=0.47]{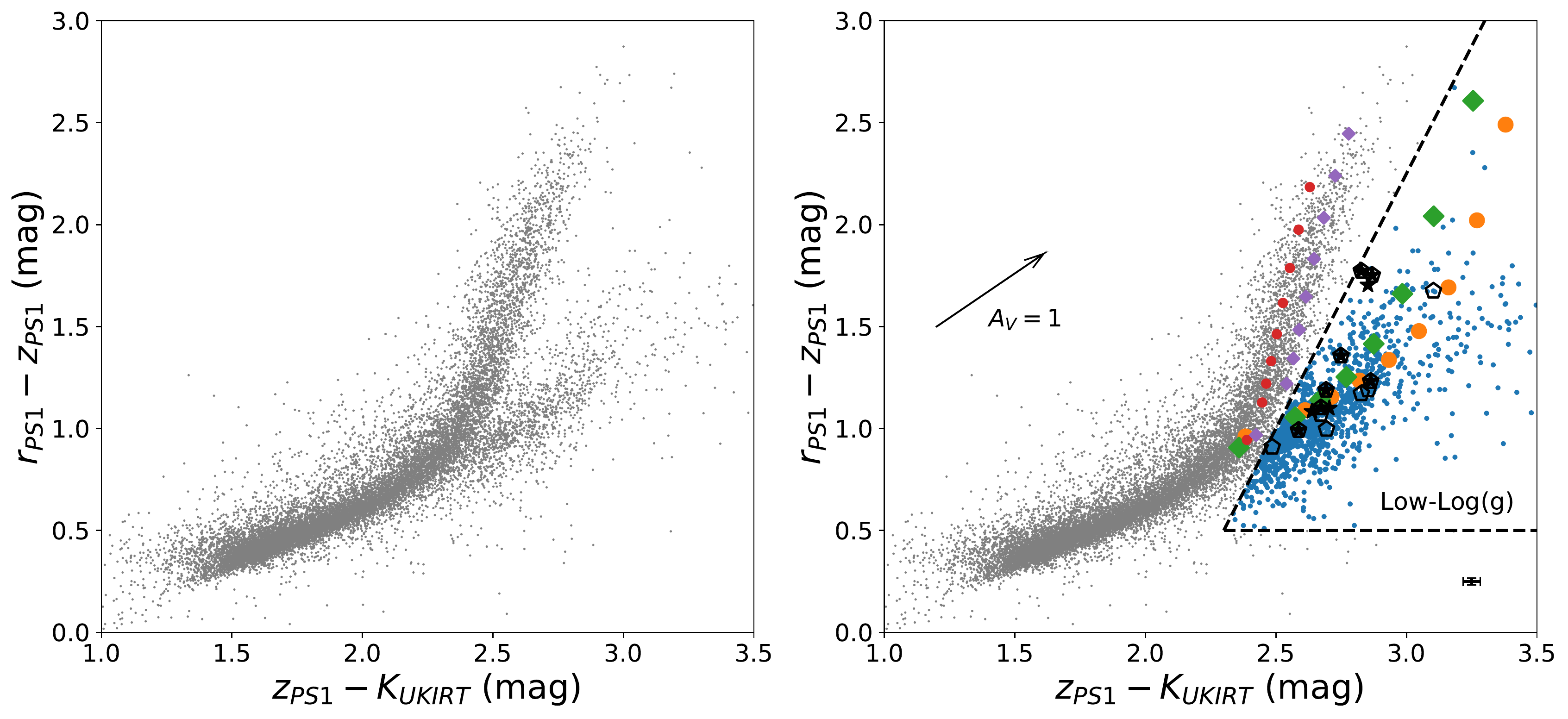}
\caption{Color-color diagrams of z-H versus r-z (upper) and z-K versus r-z (bottom). Upper left panel shows examples of both KDE (contours) and MAD (solid circles indicate the median values of z-H color in equal bins (from 0.3 to 2.6 with a step of 0.2 mag) of r-z color and error bars indicate the three times of MAD for corresponding bin), for which both of them fail to include all the spectroscopically confirmed RSGs. Left panels show the original data without the overlapping of MARCS models, while right panels show the MARCS model-selected LSG region (dashed lines which are made by eye). Small and big solid circles and diamonds represent HSG and LSG targets at different $T_{\rm eff}$ (from 3300 to 4250K) in different metallicities derived from MARCS models, respectively. Spectroscopically confirmed RSGs from \citet{Levesque2012} and \citet{Patrick2015} are shown as open pentagon and solid stars, respectively. A reddening vector of $A_{\rm V}=1.0$~mag is shown as a reference (same below). Error bars at bottom right of right panels indicate the median error of each color.
\label{ngc6822_ccd_ps1_ukirt}}
\end{figure*}

\begin{figure}
\center
\includegraphics[scale=0.47]{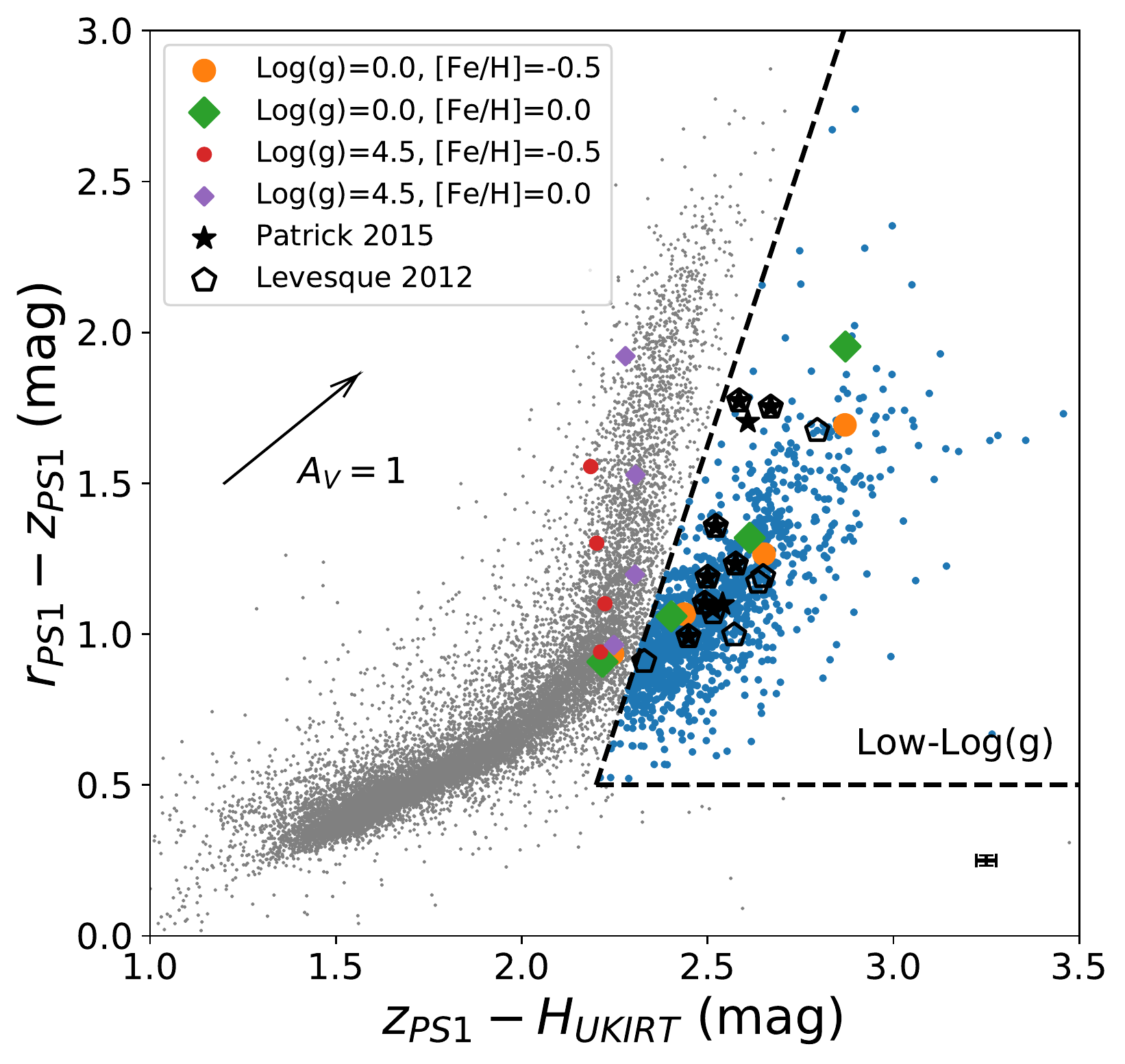}
\caption{Same as Figure~\ref{ngc6822_ccd_ps1_ukirt} but for models from ATLAS9 at $T_{\rm eff}=3500, 3750, 4000, 4250$ K.
\label{ngc6822_ccd_ps1_ukirt_atlas9}}
\end{figure}

To test this assumption, we crossmatched data between PS1 and \citet{Sibbons2012} (UKIRT) with a search radius of $1''$, which resulted in 15,508 targets. The MARCS spectra ($T_{\rm eff}=3300, 3400, 3500, 3600, 3700, 3800, 3900, 4000, 4250$ K; $Log~g=0.0, 4.5$; $Z=-0.75, 0.0$) were resampled and convolved with filters from both PS1 and UKIRT. The magnitudes and colors were all on the WFCAM instrumental system \citep{Hodgkin2009}. We investigated \textit{ALL} color combinations from PS1 and UKIRT, finding that two of them (z-H versus r-z (rzH) and z-K versus r-z (rzK)) successfully separated out foreground HSG dwarfs and background LSG targets as shown in Figure~\ref{ngc6822_ccd_ps1_ukirt} (we noted that, in contrast, the color combinations involved with J-filter were not as good as H and K). From the diagram, it can be seen that the expected color-color distributions of LSG targets ($Log~g\sim0.0$) are coincident with the observational data after applying appropriate extinction (we adopted $A_{\rm V}=1.0$~mag as a combination of both foreground and intrinsic extinction with $E(r-z)=0.356$, $E(z-h)=0.356$, and $E(z-K)=0.409$ for NGC 6822; \citealt{Wang2019}). Initially, we used several statistics methods (e.g., kernel density estimation (KDE), median absolute deviation (MAD), histogram, and so on) trying to quantitatively separate the HSG and LSG regions. However, all of the methods failed to include \textit{ALL} the spectroscopically confirmed RSGs from \citet{Levesque2012} and \citet{Patrick2015}. The upper left panel of Figure~\ref{ngc6822_ccd_ps1_ukirt} shows examples of both KDE (contours) and MAD (solid circles with error bars). For KDE, it does a relatively good job to separate LSG from HSG targets, but fails to include a few of spectroscopically confirmed RSGs located right in the middle of HSG and LSG populations. For MAD, where solid circles indicate the median values of z-H color in equal bins (from 0.3 to 2.6 with a step of 0.2 mag) of r-z color and error bars indicate the three times of MAD of z-H color for corresponding bins, it properly represents the HSG population for the most part but fails at the junction of the HSG and LSG populations. Thus, We eventually defined the LSG region (dashed lines) by eye as, 
\begin{equation}
(r_{PS1}-z_{PS1})<3.75\times(z_{PS1}-H_{UKIRT})-7.75,
\end{equation}
or
\begin{equation}
(r_{PS1}-z_{PS1})<2.5\times(z_{PS1}-K_{UKIRT})-5.25,
\end{equation}
and for both CCDs, 
\begin{equation}
(r_{PS1}-z_{PS1})\geq0.5.
\end{equation}
Meanwhile, we also showed the datasets with and without the overlapping of models, which clearly indicated that our boundaries were appropriate. Spectroscopically confirmed RSGs from \citet{Levesque2012} and \citet{Patrick2015} are shown as open pentagon and solid stars, respectively, which are all located within the LSG region. Moreover, further investigation by using synthetic photometry from ATLAS9 models\footnote{http://svo2.cab.inta-csic.es/theory/newov2/syph.php} ($T_{\rm eff}=3500, 3750, 4000, 4250$ K; $Log~g=0.0, 4.5$; $Z=-0.5, 0.0$; \citealt{Castelli2003}) also gave similar results as shown in Figure~\ref{ngc6822_ccd_ps1_ukirt_atlas9}. In total, we selected 1,276 and 1,194 LSG targets from rzH and rzK CCDs with 1,051 targets in common, respectively.

\begin{figure*}
\center
\includegraphics[scale=0.5]{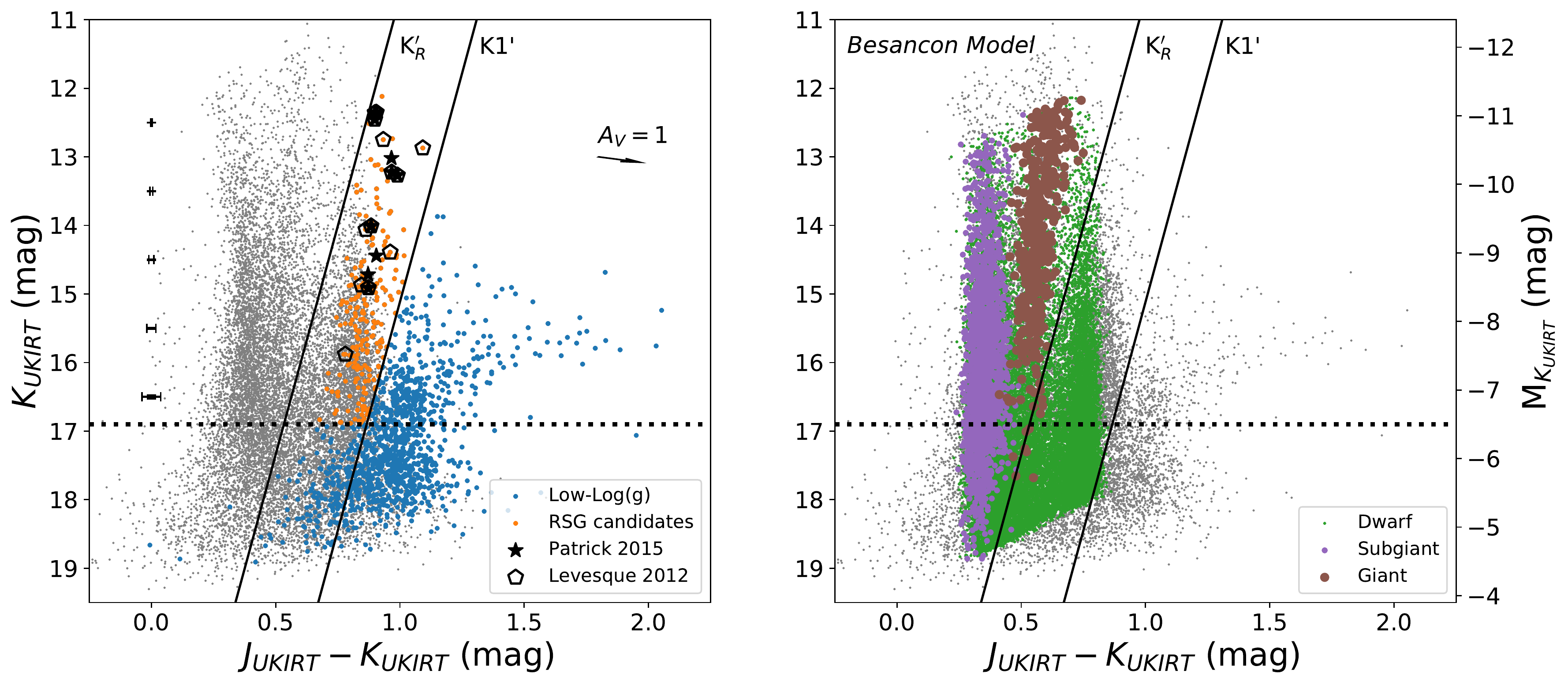}
\caption{Color-magnitude diagrams of K versus $J-K$ for observational data (left) and the Besan\c{c}on models along the line of sight of NGC 6822 (right). From the left panel, 227 RSG candidates are selected based on a slightly modified selection criteria indicated by $K1'$ and $K'_{\rm R}$ lines (221 based on rzH, 200 based on rzK, and 194 in common). Error bars at left indicate the median error of each magnitude interval. The same selection criteria in the right panel suggest that the RSG sample is almost free from contamination of Galactic giants.
\label{ngc6822_cmd_ps1_ukirt}}
\end{figure*}

After collecting LSG targets, the next step was to separate RSG candidates from the rest of the LSG candidates as determined by the ``H-bump'' method, since the majority of the LSG targets would be red giants (RGs) or asymptotic giant branch stars (AGBs). This was relatively easy as the selected targets were assumed to be located in the NGC6822 with the same distance, so that the RSGs would be brighter than RGs and AGBs. The left panel of Figure~\ref{ngc6822_cmd_ps1_ukirt} shows the color-magnitude diagram (CMD) of $K_{\rm S}$ versus $J-K_{\rm S}$. Previously, we identified RSG population in the SMC by using theoretical distribution in NIR CMD, where $K1$, $K2$, and $K_{\rm R}$ lines were used to separate out the Oxygen-rich AGB, Carbon-rich AGB, and RSG populations (see more details in \citealt{Cioni2006, Yang2011, Boyer2012, Yang2012, Yang2018, Yang2019, Yang2020a}). Here we applied a slightly modified selection criteria based on previous studies. It is important to notice that, as indicated by \citet{Yang2020a}, there is a continuum with similarity and overlapping between RSGs and AGBs in spectra, lightcurves, and CMDs. Thus, the clear separation between them is still a pending issue and cannot be accurately parameterized. The intercept of $K1'$ line, which separated RSGs from AGBs, was changed depending on the morphology of the CMD and distance modulus. The nearly vertical branch on the CMD was most likely the RSG population, while the AGB population stretched towards the red end. The $K'_{\rm R}$ line indicated the blue boundary ($\Delta(J-K)=0.333$ mag from $K1'$) of the RSG population. In order to avoid contamination from RGs and AGBs, the lower limit of the magnitude of RSG population (horizontal dotted line) was set to $M_{\rm K}=-6.5$ mag ($K=16.9$ mag; $\sim$0.5 mag above the K-band tip of red giant branch ($K-TRGB=17.36\pm0.04$ mag); \citealt{Hirschauer2020}). However, inevitably, we might lose some low-luminosity RSGs located between our lower limit and the K-TRGB. In addition to the Galactic dwarfs, another source of contamination are RGs in the Galactic halo. To address this issue, we used Besan\c{c}on models of the MW \citep{Robin2003} to predict the location of foreground stars along the line of sight of NGC 6822 as shown in the right panel of Figure~\ref{ngc6822_cmd_ps1_ukirt}. It can be seen from both panels that our selection largely avoids the majority of Galactic giants contamination, except in the faint end close to the K-TRGB as is expected. In total, we selected 227 RSG candidates based on the combination of CCDs and CMDs from PS1 and UKIRT, where 221 based on rzH, 200 based on rzK, and 194 ($\sim$85\%) in common.

As different NIR surveys cover different parts of NGC 6822 with different sensitivities, depths, and quality cuts, they may complement each other. Thus, we also crossmatched PS1 with other NIR data. The crossmatching between PS1 and VHS/IRSF/HAWK-I resulted in 11515, 2207, and 958 targets, respectively. We selected 211 RSG candidates based on PS1 versus VHS (Figure~\ref{ngc6822_ccd_cmd_ps1_vhs}; no H-band data from VHS), 183 RSG candidates based on PS1 versus IRSF (Figure~\ref{ngc6822_ccd_cmd_ps1_irsf}; 176 based on rzH, 178 based on rzKs, and 171 in common), and 88 RSG candidates based on PS1 versus HAWK-I (Figure~\ref{ngc6822_ccd_cmd_ps1_hwaki}; no H-band data from HAWK-I), respectively. We present all the CCDs and CMDs in the Appendix~\ref{appendix2}. Meanwhile, despite the slight differences in the filter sets (e.g., HK$_{(\rm S)}$ between UKIRT, VISTA, IRSF, and HAWK-I), they all give similar results. It means that any sort of HK$_{(\rm S)}$ filters can be used in future studies. We only compared NIR filters since the differences are much smaller between optical filters than the NIR filters. After removing the duplications, in total, there are 323 RSG candidates based on PS1 and NIR data.

\begin{figure*}
\center
\includegraphics[scale=0.5]{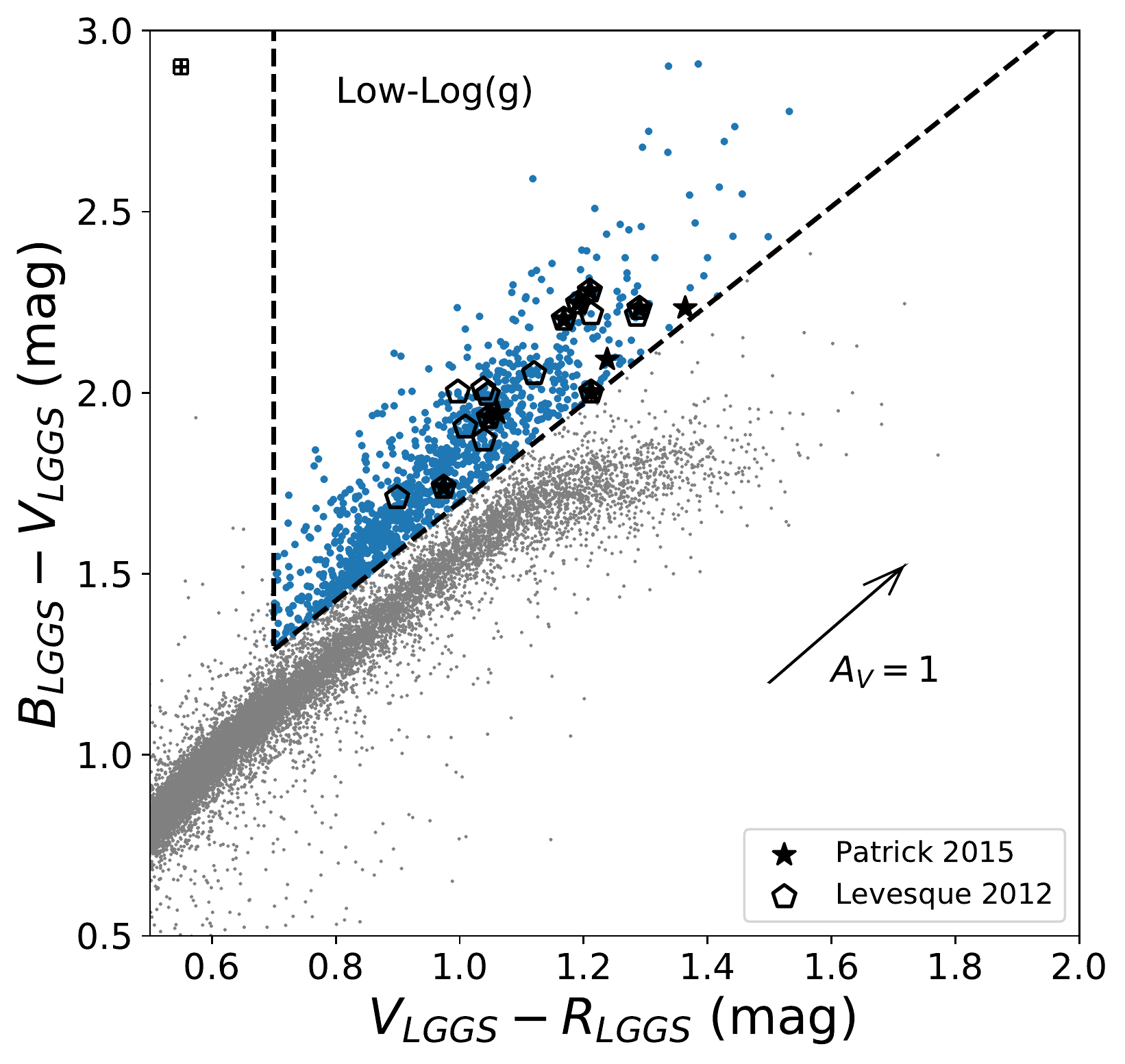}
\includegraphics[scale=0.5]{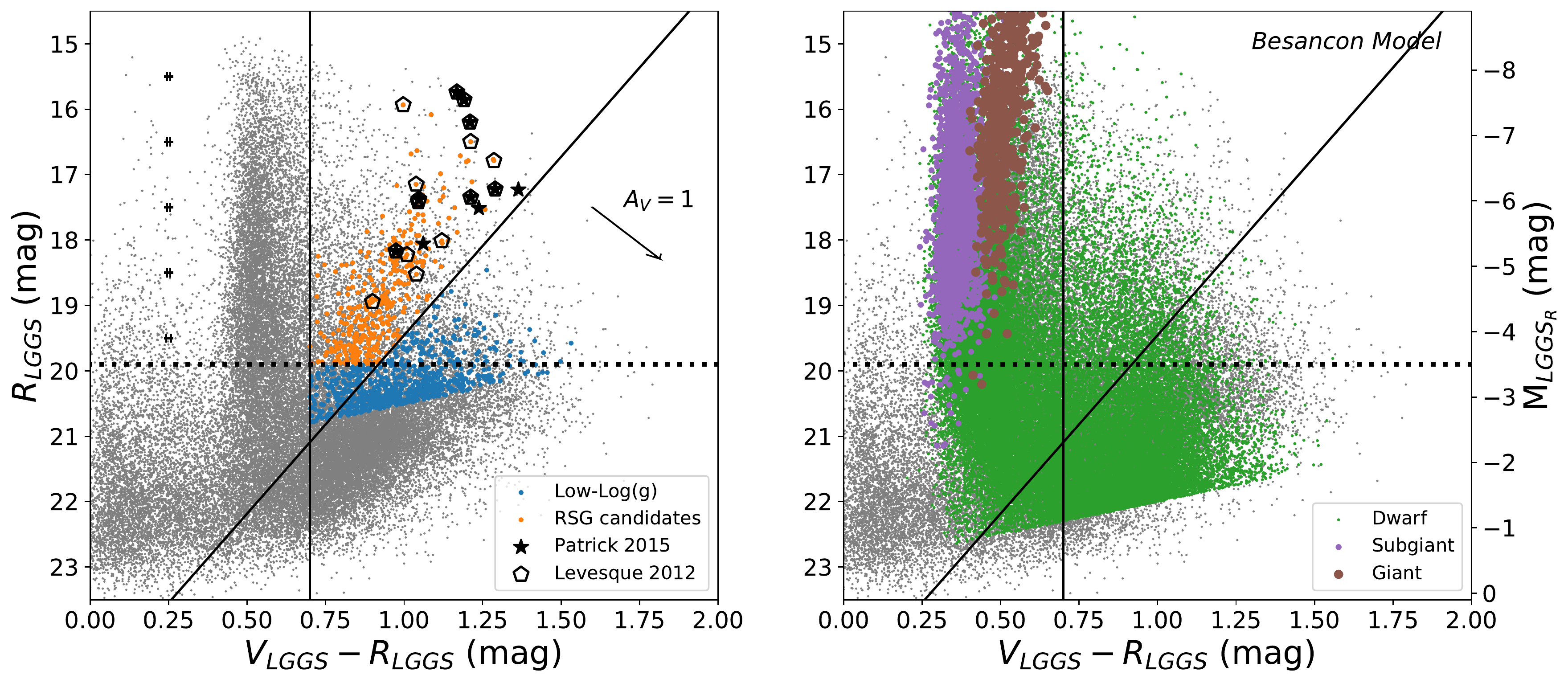}
\caption{Same as Figure~\ref{ngc6822_ccd_ps1_ukirt} and \ref{ngc6822_cmd_ps1_ukirt}, but for LGGS data. 358 RSG candidates are selected.
\label{ngc6822_ccd_cmd_lggs}}
\end{figure*}

In addition to the ``H-bump'' method we adopted to identify RSG candidates, we also applied the traditional BVR method as a comparison and/or supplement. Figure~\ref{ngc6822_ccd_cmd_lggs} shows the BVR CCD and R versus V-R CMD. Here we used slightly aggressive cuts (a cut of $V\leq21.5$ mag was applied in order to reduce the contamination) in the CCD to select as much as possible RSG candidates, since the sample would be further constrained by the CMD. Meanwhile, the cuts in the CMD were made by eye in order to select the RSG branch based on its morphology, which due to the ambiguous boundary between RSGs and AGBs as mentioned before. The lower limit of the magnitude of RSG population was set to $M_{\rm R}=-3.5$ mag ($R=19.9$ mag; about half mag above the $R-TRGB\approx20.6$ mag, which was roughly determined by a saddle point method; \citealt{Ren2020}). Similar simulation from the Besan\c{c}on models indicated that the selection of RSG candidates was also largely free from foreground contamination of giants. In total, we selected 358 RSG candidates based on LGGS data.

\section{Constraining the red supergiant candidates}

\begin{figure*}
\center
\includegraphics[scale=0.5]{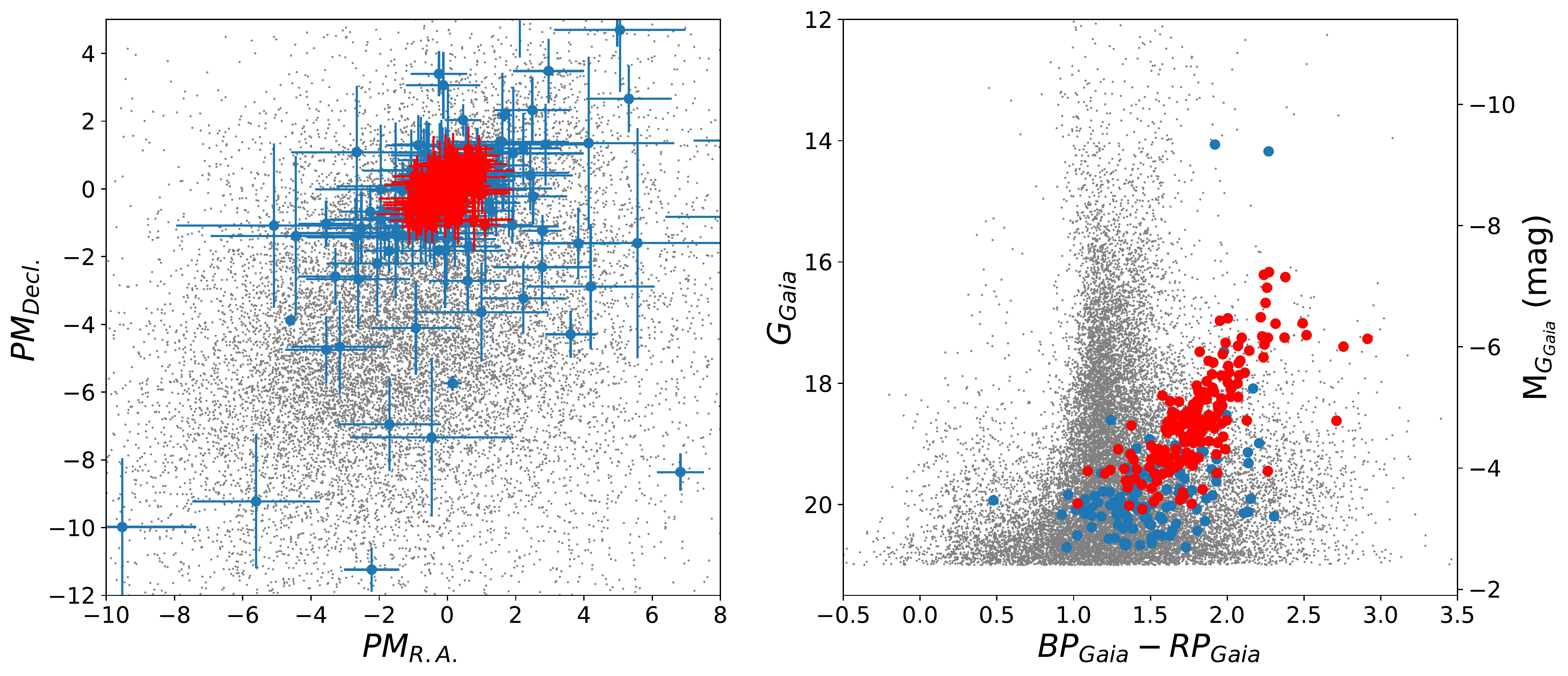}
\includegraphics[scale=0.5]{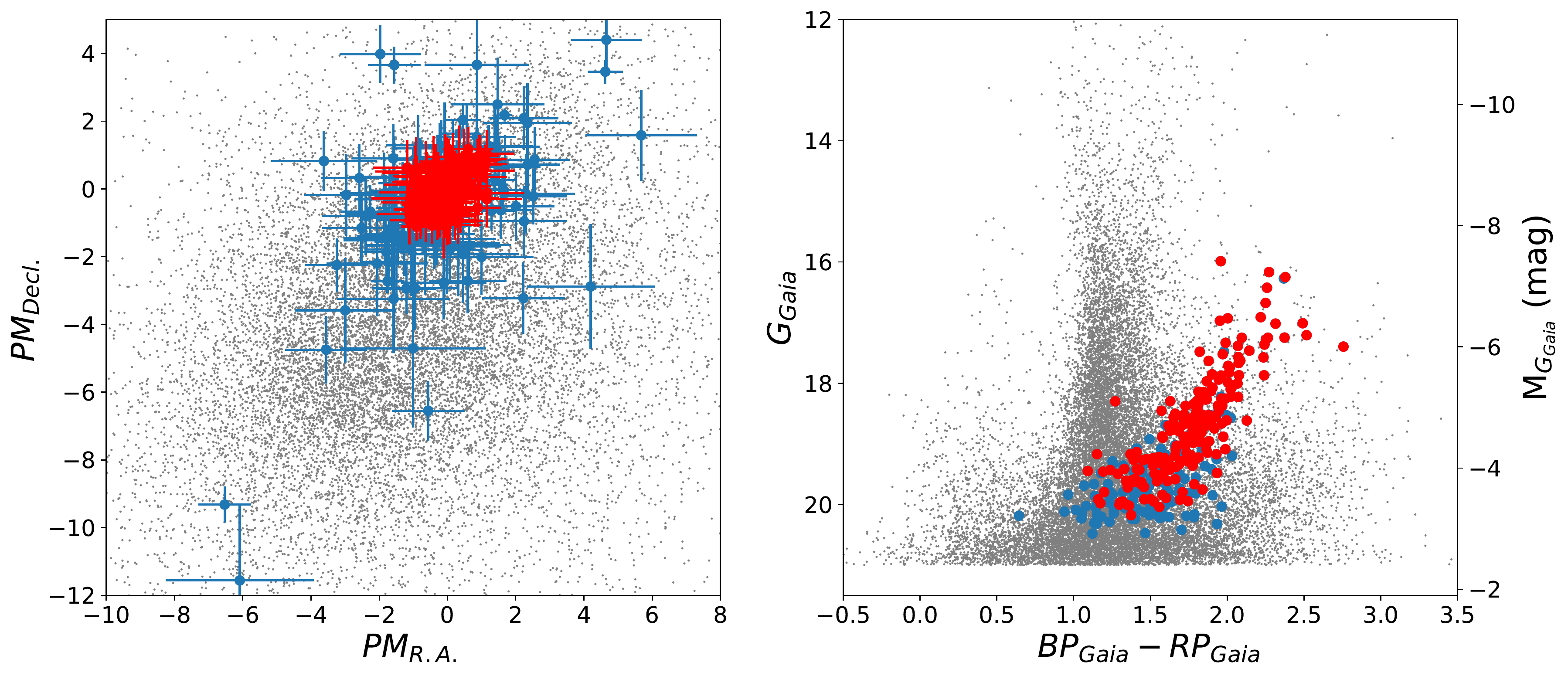}
\caption{Evaluation of Gaia astrometric solution (left) and CMDs (right) for RSG candidates from ``H-bump'' (top) and BVR method (bottom), respectively. The final selected RSG candidates (red) are clustering around zero in PMs, while outliers (blue) are mainly in the faint and blue end of the CMDs.
\label{ngc6822_candi_gaia}}
\end{figure*}

Further constraining the RSG candidates was carried out by using the Gaia astrometric solution. The RSG candidates from both methods were crossmatched with Gaia DR2 with a search radius of $1''$, which resulted in 309 targets for ``H-bump'' and 314 targets for BVR method, respectively. Targets without Gaia data were not considered for further analysis. For the members of NGC 6822, the proper motions (PMs) and parallaxes of targets are supposed to be around zero. In that sense, we constrained the Gaia astrometric solution and color of final RSG candidates as listed in Table~\ref{gaia_constraint}. The constraints are justified based on the typical errors of $\sim$1.2 mas/yr for the PMs and $\sim$0.7 mas for the parallaxes at the faint end of Gaia magnitude, respectively \citep{Lindegren2018}. Notice that, we did not apply the same statistical method in \citet{Yang2019}, for which we used Gaussian profiles to fit Gaia astrometric solution in order to select the proper members of the SMC. This is mainly due to two reasons. Firstly, for the SMC (or any other nearby galaxies with large angular size and relatively high Galactic latitude), the dominant population in the FoV is the members of the target galaxy. However, for the case of NGC 6822 with relatively small angular size and low Galactic latitude (plus it is more distant than the SMC), the dominant population is no longer its members but the foreground stars. Secondly, a large fraction of the members of NGC 6822 do not have proper astrometric measurements due to either faintness or crowding (it is estimated that the members of NGC 6822 only comprise around 10\% population of measurable astrometric solution in the FoV based on Figure~\ref{ngc6822_candi_gaia}). In that sense, the statistics of astrometric solution in the FoV will be highly biased due to the large contamination of foreground stars. 

Figure~\ref{ngc6822_candi_gaia} shows the constraints on the Gaia astrometric solution and CMDs for both ``H-bump'' (top) and BVR (bottom) method. There are 181 and 193 targets selected for ``H-bump'' and BVR method, respectively. The selected targets (red) are concentrated around PMs of zero, and are mainly limited to $G_{\rm Gaia}\approx20.0$ mag due to the relatively strict constraints of Gaia astrometric solution as mentioned above. Meanwhile, the outliers (blue), as expected, are mostly located in the faint end on the CMDs, due to large PMs (parallax) and/or large PM (parallax) errors. We note that among the blue outlier points, some genuine RSGs are expected to exist. One may argue that faint targets without Gaia astrometric solution could also belong to NGC 6822 due to their distances. However, another possibility is that they are the distant Galactic red dwarfs, especially since NGC 6822 is in relatively low Galactic latitude. Notice that, the foreground contamination can be mitigated by taking conservative cuts in the CCD, CMD, and astrometric solution. However, it will also cause target loss in the faint and blue end. The percentages of selected targets in both methods are similar as $\sim$59\% (181/309) for ``H-bump'' and $\sim$61\% (193/314) for BVR method, which indicates that the accuracy of two methods is comparable. In total, there are 234 RSG candidates after combining targets from both methods with 140 ($\sim$60\%) of them in common. The full information about the 234 candidates is listed in Table~\ref{sample}. We note that there is a large difference for the number of RSG candidates between NGC 6822 and the SMC (1,239 candidates; \citealt{Yang2020a}), even though they are similar in sizes and metallicities. The large number of RSG candidates in the SMC is most likely due to the interaction between the MW, the LMC, and the SMC, which triggers multiple peaks of star formation in the past few hundreds million years (an underlying constant SFR of $\sim$0.1 M$_\sun$/yr with superposed episodes of enhanced star formation at 2–3 Gyr, 400 Myr, and 60 Myr; \citealt{Yoshizawa2003, Harris2004, Harris2009, Indu2011}). Meanwhile, since NGC 6822 is an isolated dIrr, its star formation rate is relatively low and constant ($\sim$0.014 M$_\sun$/yr over the past 100 Myr, and $\sim$0.01 M$_\sun$/yr in the most recent 10 Myr \citealt{Wyder2001, Efremova2011, Fusco2014}), resulting in a small number of RSG candidates.

\begin{figure*}
\center
\includegraphics[scale=0.33]{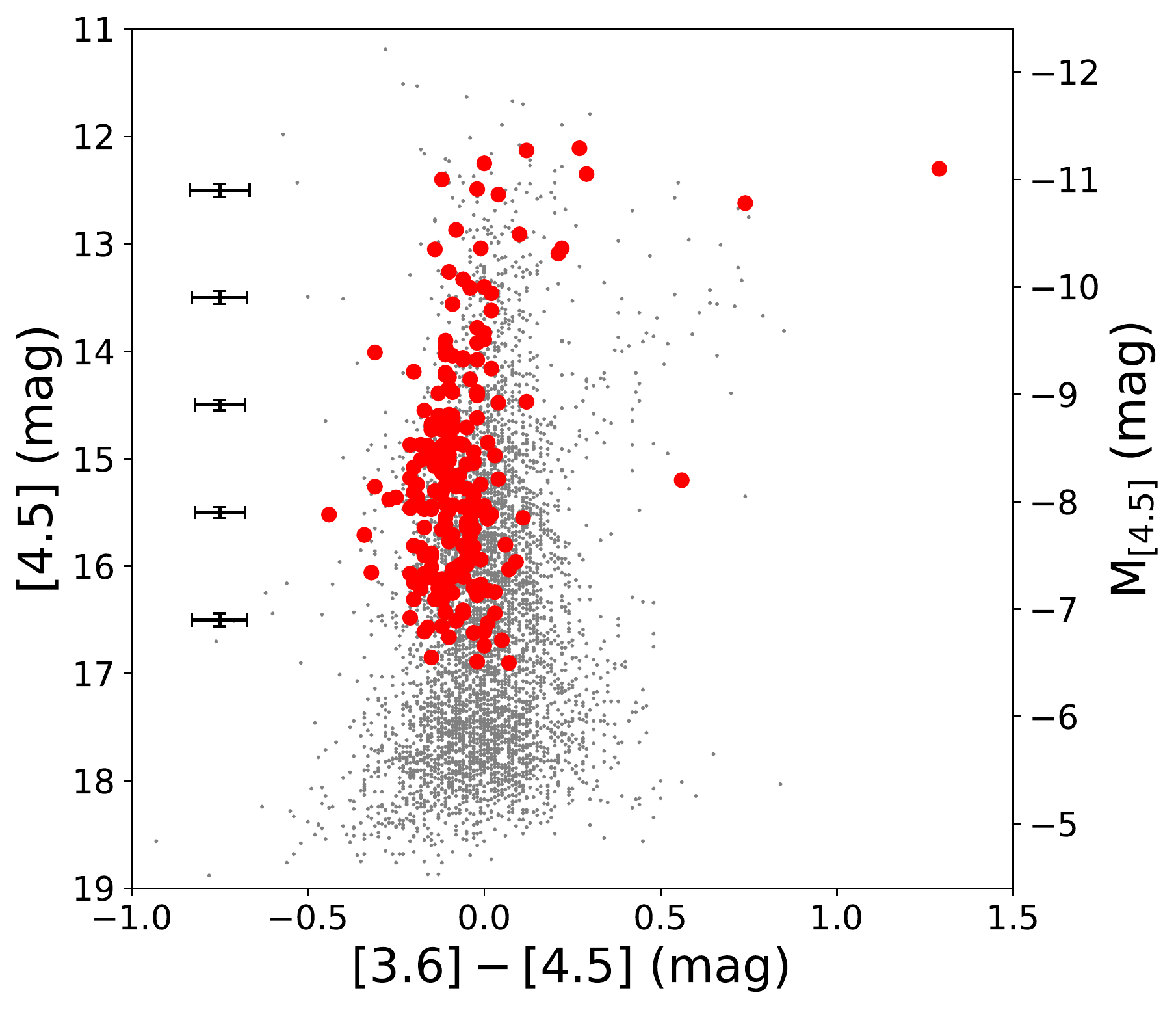}
\includegraphics[scale=0.33]{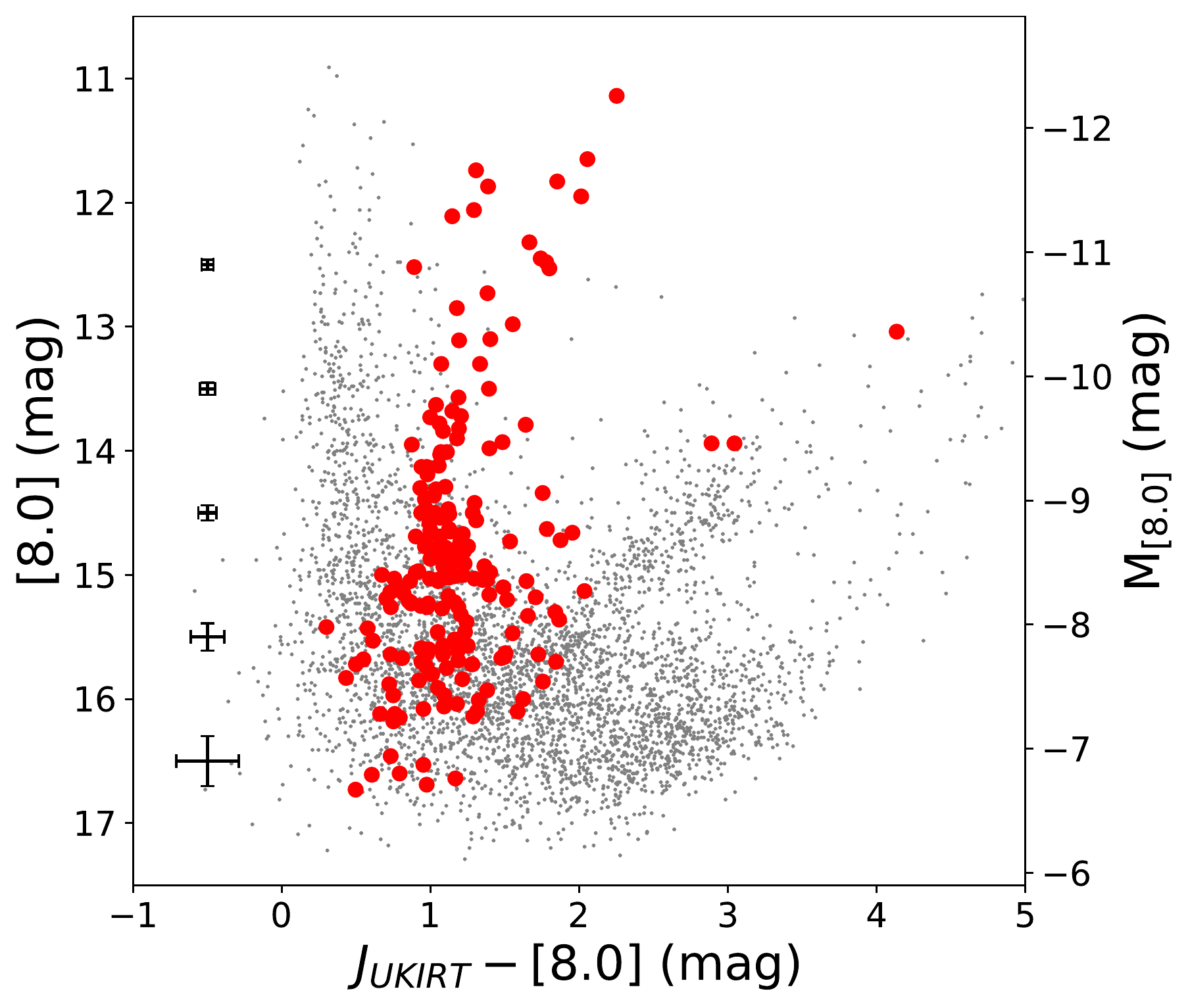}
\includegraphics[scale=0.33]{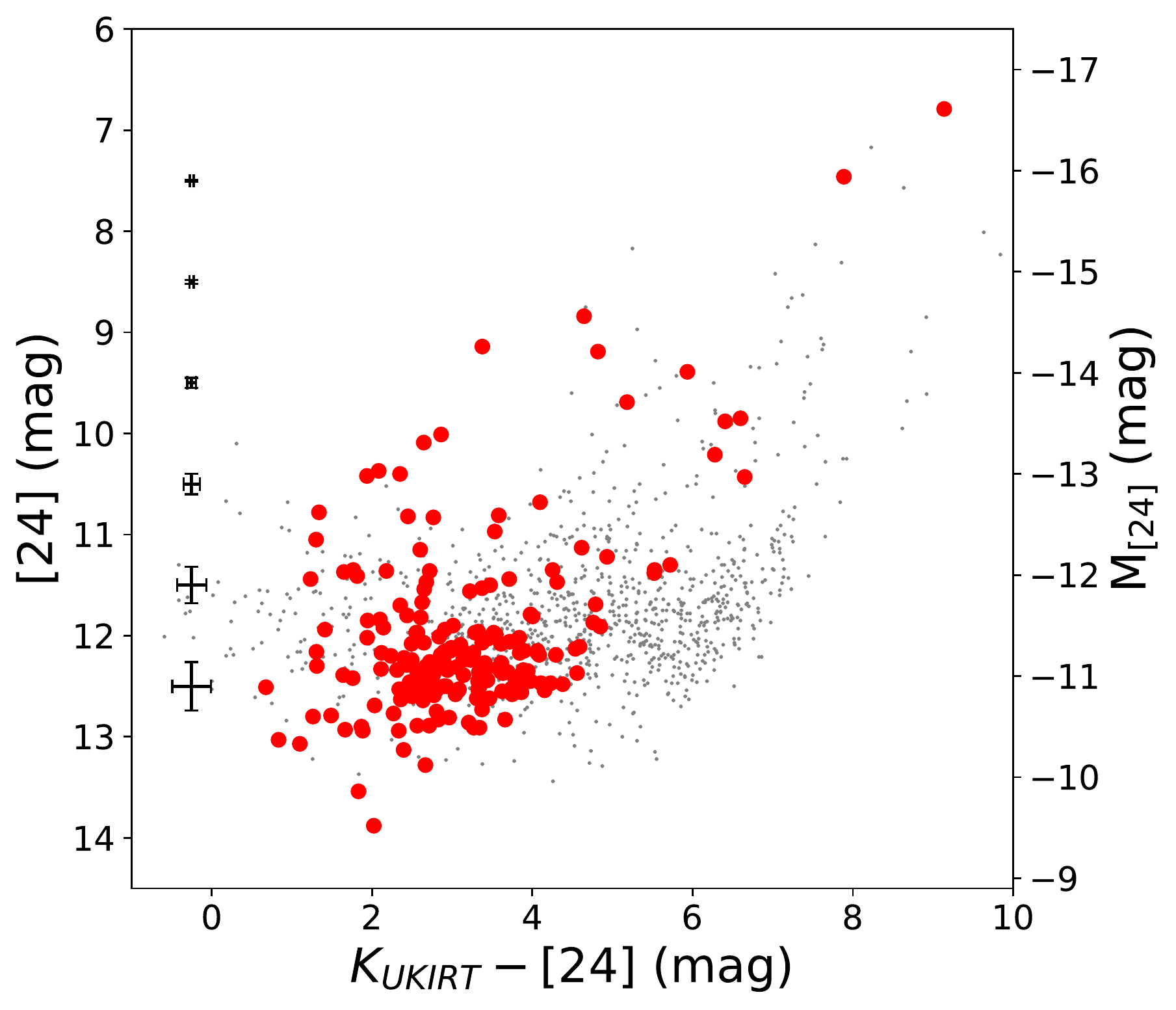}
\caption{Mid-infrared color-magnitude diagrams of [4.5] versus $[3.6]-[4.5]$ (left), [8.0] versus $J-[8.0]$ (middle), and [24] versus $K-[24]$ (right). The RSG candidates are located in the expected position on each diagram as $[3.6]-[4.5]\lesssim0$ and $J-[8.0]\approx1.0$. The large spreading in $K-[24]$ is most likely due to the low sensitivity in [24]-band. 
\label{ngc6822_candi_mir_cmds}}
\end{figure*}

\begin{figure}
\center
\includegraphics[scale=0.5]{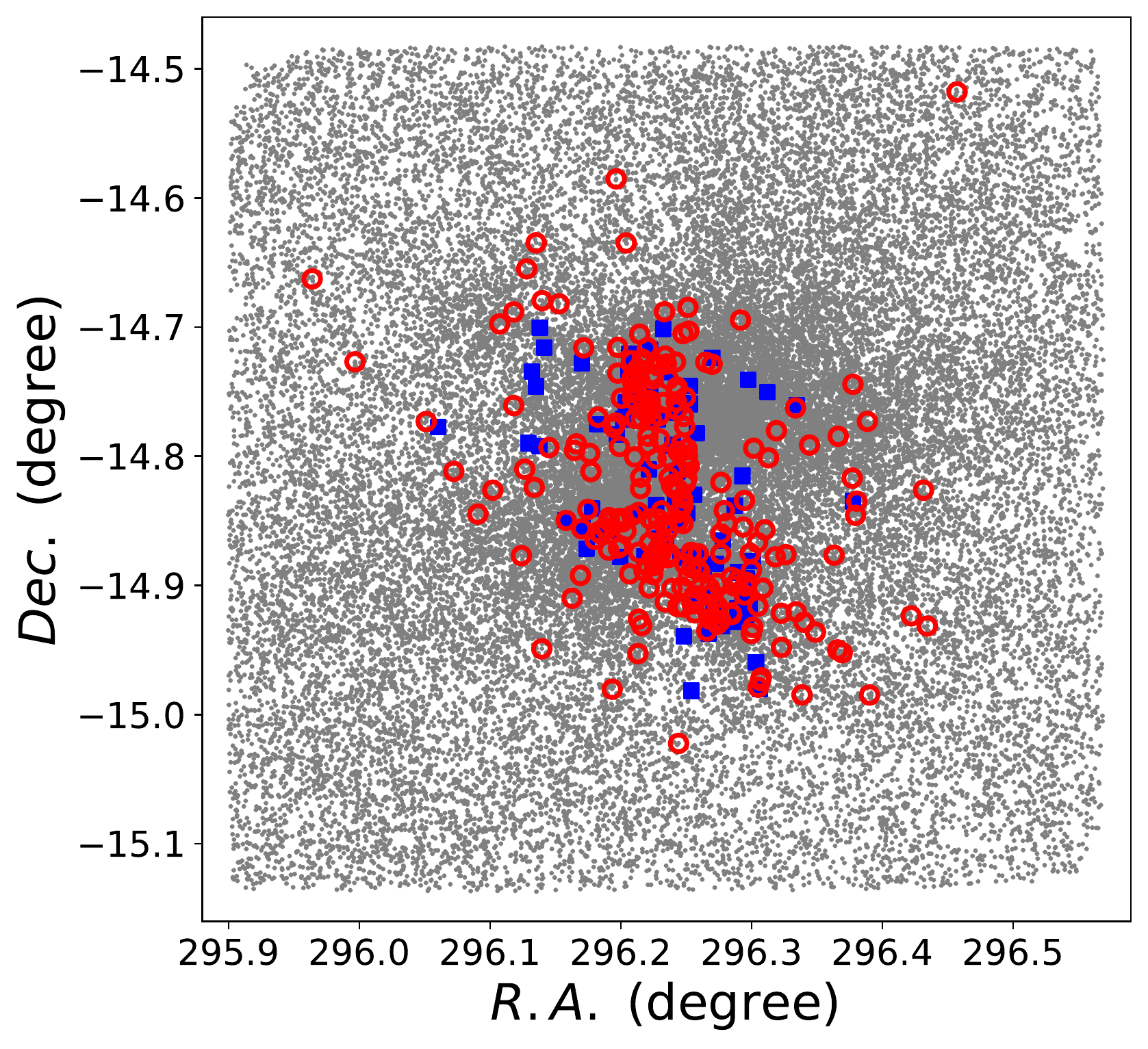}
\caption{Spatial distribution of RSG candidates (red open circles). It follows nicely the distribution of FUV-selected star formation regions (blue solid squares; \citealt{Efremova2011}).
\label{ngc6822_candi_spatial}}
\end{figure}

In addition to previous optical and NIR CMDs, we also plotted MIR CMDs dominated by the dust emission to further evaluate the sample. Figure~\ref{ngc6822_candi_mir_cmds} show CMDs of [4.5] versus $[3.6]-[4.5]$, [8.0] versus $J-[8.0]$, and [24] versus $K-[24]$. Background targets are the result of crossmatching ($1''$ search radius) between Spitzer and UKIRT data, without upper limit and extended targets ($\lvert [3.6]_{\rm PSF}-[3.6]_{\rm Aper} \rvert\leq0.1$). It can be seen that the selected RSG candidates are all in the expected locations (see Figure 5 of \citealt{Yang2020a}), with blue color ($\lesssim0$) in $[3.6]-[4.5]$ and red color ($\sim1.0$) in $J-[8.0]$. The spreading in $K-[24]$ is most likely due to the low sensitivity in [24]-band. Finally, Figure~\ref{ngc6822_candi_spatial} shows the spatial distribution of RSG candidates, for which it follows well the far-ultraviolet (FUV) selected star formation regions from \citet{Efremova2011}. This also indicates that our selection of RSG candidates is reasonable and reliable. The comparison between the spatial distribution of our sample and the result from \citet{Hirschauer2020} (see Figure 12 of their paper) may indicate that, our sample is purer with less contamination from low mass and foreground stars, since the distribution of RSGs from \citet{Hirschauer2020} is less compact.

Finally, we would like to point out that, our method can also be used to identify other LSG targets like RGs and AGBs as shown in Figure~\ref{ngc6822_cmd_ps1_ukirt}. Combining the high-resolution and deep photometry from most recent large-scale ground-based surveys, for example, VHS, UKIRT Hemisphere Survey (UHS; \citealt{Dye2018}), UKIRT Infrared Deep Sky Survey (UKIDSS; \citealt{Lawrence2007}), PS1, Legacy Survey \citep{Dey2019}, using ``H-bump'' to identify RSGs with non-Johnson filter bands can be largely applied to most of the nearby galaxies. A good example of our method has already been shown in \citet{Ren2020}. Moreover, the future ground and space facilities, like Chinese Space Station Telescope (CSST), Vera C. Rubin Observatory (LSST), Euclid, Nancy Grace Roman Space Telescope (NGRST), James Webb Space Telescope (JWST), may promote further application beyond the LG, uncovering unprecedented populations of RSGs in a vast range of different galactic environments.

\section{Summary}

We present a novel method to identify RSG candidates based on their 1.6 $\mu$m ``H-bump'', for which a case study is carried out for NGC 6822, due to its rich reservoir in multiwavelength data. 32 bands of data were collected ranging from optical to MIR, derived from Gaia, PS1, LGGS, VHS, UKIRT, IRSF, HAWK-I, Spitzer, and WISE.

The theoretical spectra of LSG and HSG targets from MARCS were compared, which indicated a striking difference (around 1.6 $\mu$m ``H-bump'') between the two populations. To take advantage of this difference, we crossmatched optical (PS1) and NIR surveys to identify efficient color combinations that separate LSG (mostly foreground dwarfs) and HSG targets (mainly background RGs, AGBs, and RSGs). The best results are provided by rzH and rzK. Moreover, synthetic photometry derived from ATLAS9 models also gave similar results. Further separating RSG candidates from the rest was done by using semi-empirical criteria on NIR CMDs, where both the theoretic cuts and morphology of RSG population were considered. Meanwhile, simulating foreground stars along the line of sight of NGC 6822 with Besan\c{c}on models also indicated that, our selection criteria were largely free from the contamination of Galactic giants. In total, there are 323 RSG candidates based on PS1 and NIR (UKIRT, VHS, IRSF, HAWK-I) data. To test our approach we also used the traditional BVR method. We applied a slightly aggressive cut in order to select as many RSG candidates as possible, which resulted in 358 targets.

After selecting RSG candidates based on CCDs and CMDs, Gaia astrometric solution was used to further constrain the sample. We applied relatively strict criteria to Gaia PMs and parallaxes, where 181 and 193 targets were selected for ``H-bump'' and BVR method, respectively. The accuracy of the two methods is comparable as $\sim$60\%. In total, there are 234 RSG candidates after combining targets from both methods with 140 ($\sim$60\%) of them in common. 

Further evaluation of the RSG sample was done by comparing its location on different MIR CMDs and spatial distribution. The RSG candidates are found in the expected locations on the MIR CMDs with $[3.6]-[4.5]\lesssim0$ and $J-[8.0]\approx1.0$. Its spatial distribution coincides with the FUV-selected star formation regions, strengthening our belief that this selection approach is reasonable and reliable. 

Finally, we indicated that our method also can be used to identify other LSG targets like RGs and AGBs, as well as applied to most of the nearby galaxies by utilizing the recent large-scale ground-based survey. Future ground and space facilities may promote its application beyond the LG, uncovering unprecedented populations of RSGs in a vast range of different galactic environments.

\begin{acknowledgements}

We would like to thank the anonymous referee for many constructive comments and suggestions. This study has received funding from the European Research Council (ERC) under the European Union's Horizon 2020 research and innovation programme (grant agreement number 772086). B.W.J and J.G. gratefully acknowledge support from the National Natural Science Foundation of China (Grant No.11533002 and U1631104). 

This work is based in part on observations made with the Spitzer Space Telescope, which is operated by the Jet Propulsion Laboratory, California Institute of Technology under a contract with NASA. This publication makes use of data products from the Wide-field Infrared Survey Explorer, which is a joint project of the University of California, Los Angeles, and the Jet Propulsion Laboratory/California Institute of Technology. It is funded by the National Aeronautics and Space Administration. This research has made use of the NASA/IPAC Infrared Science Archive, which is operated by the Jet Propulsion Laboratory, California Institute of Technology, under contract with the National Aeronautics and Space Administration.

This work has made use of data from the European Space Agency (ESA) mission {\it Gaia} (\url{https://www.cosmos.esa.int/gaia}), processed by the {\it Gaia} Data Processing and Analysis Consortium (DPAC, \url{https://www.cosmos.esa.int/web/gaia/dpac/consortium}). Funding for the DPAC has been provided by national institutions, in particular the institutions participating in the {\it Gaia} Multilateral Agreement.

This research has made use of the SIMBAD database and VizieR catalog access tool, operated at CDS, Strasbourg, France, and the Tool for OPerations on Catalogues And Tables (TOPCAT; \citealt{Taylor2005}).

This research has made use of the Spanish Virtual Observatory (http://svo.cab.inta-csic.es) supported from the Spanish MICINN/FEDER through grant AyA2017-84089

\end{acknowledgements}

\begin{table*}
\caption{Constraints of Gaia astrometric solution and color for the final RSG sample} 
\label{gaia_constraint}
\centering
\begin{tabular}{ccccccccc}
\toprule\toprule
Parameters & Constraints\\
\midrule 
$PM_{\rm R.A.}$             & $-1.2\leq$ and $\leq 1.2~mas/yr$ \\
$PM_{\rm Decl.}$            & $-1.2\leq$ and $\leq 1.2~mas/yr$ \\
parallax                    & $-0.7\leq$ and $\leq0.7~mas$     \\
$e\_PM_{\rm R.A.~or~Decl.}$ & $\leq 1.2~mas/yr$ \\ 
$e\_parallax$               & $\leq0.7~mas$ \\ 
$BP-RP$                     & $\geq1.0~mag$ \\
\midrule 
\end{tabular}
\end{table*}

\begin{table*}
\caption{Final sample of 234 RSG candidates in NGC 6822} 
\label{sample}
\centering
\begin{tabular}{ccccccccc}
\toprule\toprule
R.A.(J215.5) & Decl.(J2015.5) & Gaia\_parallax & e\_Gaia\_parallax & ...... & [22] & e\_[22] & SNR\_[22] & Method \\
(deg)       & (deg)        & (mas)          & (mas)             & ...... & (mag) & (mag)    &          &      \\
\midrule 
295.963932  &   -14.662690 &   0.535 &  0.666 &  ...... &        &      &      &  BVR  \\
295.996679  &   -14.726901 &   0.142 &  0.373 &  ...... &        &      &      &  Both \\
296.051149  &   -14.773147 &  -0.025 &  0.245 &  ...... &  8.734 &      &  0.3 &  Both \\
296.072225  &   -14.811720 &  -0.123 &  0.569 &  ...... &  8.559 &      &  0.8 &  Both \\
296.090658  &   -14.844746 &   0.239 &  0.221 &  ...... &  8.936 &      & -0.8 &  BVR  \\
...         &   ...        &   ...   &  ...   &  ...    &  ...   &  ... &  ... &  ...  \\
\midrule 
\end{tabular}
\tablefoot{
The format of this table is a simple combination of all nine photometric catalogs in the Appendix~\ref{appendix1}, with Gaia coordinate adopted as the reference coordinate. Moreover, the discrepancy between Epoch J2000.0 and Epoch J2015.5 is very small and can be ignored (https://www.cosmos.esa.int/web/gaia/faqs\#ICRSICRF). This table is available in its entirety in CDS. A portion is shown here for guidance regarding its form and content.\\
}
\end{table*}

\begin{appendix}

\section{Photometric catalogs of nine datasets \label{appendix1}}

We list here the examples of nine photometric catalogs from Section 2. All tables are available in their entireties in CDS.

\begin{table*}
\caption{PS1 photometric catalog} 
\scriptsize
\label{ps1_cat}
\begin{tabular}{ccccccccccc}
\toprule\toprule
R.A. (J2000) & Decl. (J2000) & g      & e\_g  & SD\_g & Amp\_g & ... & y      & e\_y   & SD\_y  & Amp\_y \\
(deg)        & (deg)         & (mag)  & (mag) & (mag) & (mag)  & ... & (mag)  & (mag)  & (mag)  & (mag)  \\
\midrule 
295.899772   &   -14.845621  & 21.865 & 0.053 & 0.170 & 0.469  & ... & -999.0 & -999.0 & -999.0 & 0.000 \\
295.899801   &   -14.623661  & 19.650 & 0.016 & 0.057 & 0.137  & ... & 18.619 & 0.009  & 0.146  & 0.586 \\
295.899811   &   -14.737921  & 21.373 & 0.047 & 0.164 & 0.395  & ... & 17.817 & 0.006  & 0.093  & 0.292 \\
...          &   ...         & ...    & ...   & ...   & ...    & ... &  ...   & ...    & ...    & ...   \\
\midrule 
\end{tabular}
\tablefoot{
SD: standard deviation. Amp: amplitude (max-min). This table is available in its entirety in CDS. A portion is shown here for guidance regarding its form and content.\\
}
\end{table*}

\begin{table*}
\caption{Gaia photometric catalog} 
\scriptsize
\label{gaia_cat}
\begin{tabular}{cccccccccccc}
\toprule\toprule
R.A. (J2015.5) & Decl. (J2015.5) & parallax & e\_parallax & pmra     & e\_pmra  & pmdec    & e\_pmdec & G     & BP     & RP     & RUWE  \\
(deg)          & (deg)           & (mas)    & (mas)       & (mas/yr) & (mas/yr) & (mas/yr) & (mas/yr) &(mag)  & (mag)  & (mag)  &       \\
\midrule 
295.899801     &   -14.623661    & 0.370    & 0.313       & -1.752   & 0.509     & -4.386  & 0.450    &19.106 & 19.604 & 18.469 & 0.975 \\
295.899812     &   -14.737921    & 0.266    & 0.438       & 1.781    & 0.723     & -4.200  & 0.672    &19.422 & 20.278 & 18.190 & 1.106 \\
295.899888     &   -14.679203    & -0.371   & 0.614       & -2.644   & 0.980     & -6.851  & 0.871    &20.115 & 20.545 & 19.449 & 1.038 \\
...            &   ...           & ...      & ...         & ...      & ...       & ...     &  ...     &...    & ...    & ...    & ... \\
\midrule 
\end{tabular}
\tablefoot{
RUWE: Renormalised Unit Weight Error. The discrepancy between Epoch J2000.0 and Epoch J2015.5 is very small and can be ignored (https://www.cosmos.esa.int/web/gaia/faqs\#ICRSICRF). This table is available in its entirety in CDS. A portion is shown here for guidance regarding its form and content.\\
}
\end{table*}

\begin{table*}
\caption{LGGS photometric catalog} 
\scriptsize
\label{lggs_cat}
\begin{tabular}{ccccccccccccc}
\toprule\toprule
R.A. (J2000) & Decl. (J2000) & V    & e\_V & B-V  & e\_B-V & U-B  & e\_U-B & V-R  & e\_V-R & R-I  & e\_R-I & Spectral type \\
(deg)& (deg) & (mag)& (mag)& (mag)& (mag)  & (mag)& (mag)  & (mag)& (mag)  & (mag)& (mag)  &  \\
\midrule 
295.899750  &   -14.934861 & 21.023 & 0.028 & 1.247 & 0.056 &      &      & 0.827 & 0.041 & 0.818 & 0.03  &     \\
295.899792  &   -14.845417 & 21.733 & 0.072 & 1.126 & 0.106 &      &      & 0.578 & 0.091 & 0.779 & 0.056 &     \\
295.900042  &   -15.010917 & 21.683 & 0.038 & 1.023 & 0.081 &      &      & 0.544 & 0.064 & 0.552 & 0.052 &     \\
...         &   ...        & ...    & ...   & ...   & ...   &  ... &  ... & ...   & ...   & ...   & ...   & ... \\
\midrule 
\end{tabular}
\tablefoot{
This table is available in its entirety in CDS. A portion is shown here for guidance regarding its form and content.\\
}
\end{table*}

\begin{table*}
\caption{UKIRT photometric catalog} 
\scriptsize
\label{ukirt_cat}
\begin{tabular}{cccccccc}
\toprule\toprule
R.A. (J2000) & Decl. (J2000) & J      & e\_J  & H      & e\_H  & K      & e\_K  \\
(deg)        & (deg)         & (mag)  & (mag) & (mag)  & (mag) & (mag)  & (mag) \\
\midrule 
295.899750   &   -14.510397  & 18.873 & 0.111 & 18.584 & 0.101 & 18.196 & 0.100 \\
295.899750   &   -14.610637  & 18.853 & 0.110 & 18.453 & 0.091 & 18.253 & 0.106 \\
295.899750   &   -14.784898  & 16.839 & 0.024 & 16.196 & 0.017 & 16.024 & 0.018 \\
...          &   ...         & ...    & ...   & ...    & ...   &  ...   &  ...  \\
\midrule 
\end{tabular}
\tablefoot{
This table is available in its entirety in CDS. A portion is shown here for guidance regarding its form and content.\\
}
\end{table*}

\begin{table*}
\caption{VHS photometric catalog} 
\scriptsize
\label{VHS_cat}
\begin{tabular}{cccccccc}
\toprule\toprule
R.A. (J2000) & Decl. (J2000) & J      & e\_J  & K$_{\rm S}$ & e\_K$_{\rm S}$ \\
(deg)        & (deg)         & (mag)  & (mag) & (mag)       & (mag)          \\
\midrule 
295.899805   &   -14.737884  & 16.504 & 0.017 & 15.557      & 0.031          \\
295.899887   &   -14.679201  & 18.418 & 0.096 & 17.941      & 0.272          \\
295.899980   &   -15.041392  & 17.961 & 0.063 & 17.273      & 0.147          \\
...          &   ...         & ...    & ...   & ...         & ...            \\
\midrule 
\end{tabular}
\tablefoot{
This table is available in its entirety in CDS. A portion is shown here for guidance regarding its form and content.\\
}
\end{table*}

\begin{table*}
\caption{IRSF photometric catalog} 
\scriptsize
\label{irsf_cat}
\begin{tabular}{cccccccc}
\toprule\toprule
R.A. (J2000) & Decl. (J2000) & J      & SD\_J & H      & SD\_H & K$_{\rm S}$ & SD\_K$_{\rm S}$ \\
(deg)        & (deg)         & (mag)  & (mag) & (mag)  & (mag) & (mag)       & (mag)           \\
\midrule 
296.170610   &   -14.952340  & 17.588 & 0.087 & 17.170 & 0.044 & 17.138      & 0.078           \\
296.170620   &   -14.960280  & 18.749 & 0.136 & 17.952 & 0.057 & 17.686      & 0.069           \\
296.170760   &   -14.970890  & 17.537 & 0.033 & 17.091 & 0.040 & 17.061      & 0.078           \\
...          &   ...         & ...    & ...   & ...    & ...   &  ...        &  ...            \\
\midrule 
\end{tabular}
\tablefoot{
SD: standard deviation. This table is available in its entirety in CDS. A portion is shown here for guidance regarding its form and content.\\
}
\end{table*}

\begin{table*}
\caption{HAWK-I photometric catalog} 
\scriptsize
\label{hawki_cat}
\begin{tabular}{cccccccc}
\toprule\toprule
R.A. (J2000) & Decl. (J2000) & J      & e\_J  & K$_{\rm S}$ & e\_K$_{\rm S}$ \\
(deg)        & (deg)         & (mag)  & (mag) & (mag)       & (mag)          \\
\midrule 
296.126853   &   -14.805602  & 17.911 & 0.036 & 16.937      & 0.033          \\
296.127634   &   -14.818471  & 18.841 & 0.022 & 17.782      & 0.084          \\
296.127794   &   -14.805742  & 18.995 & 0.020 & 18.080      & 0.068          \\
...          &   ...         & ...    & ...   & ...         & ...            \\
\midrule 
\end{tabular}
\tablefoot{
This table is available in its entirety in CDS. A portion is shown here for guidance regarding its form and content.\\
}
\end{table*}

\begin{table*}
\caption{Spitzer photometric catalog} 
\scriptsize
\label{spitzer_cat}
\begin{tabular}{ccccccccccc}
\toprule\toprule
R.A. (J2000) & Decl. (J2000) & [3.6]  & e\_[3.6] & d[3.6] & ... & I\_[5.8] & [5.8]  & e\_[5.8] & d[5.8] & ... \\
(deg)        & (deg)         & (mag)  & (mag)    & (mag)  & ... & (mag)    & (mag)  & (mag)    & (mag)   & ... \\
\midrule 
296.024140   &   -14.678080  & 18.680 & 0.160    & 0.000  & ... &          & 16.400 & 0.160    & 0.000  & ... \\
296.025180   &   -14.682550  & 19.070 & 0.140    & 0.420  & ... & >        & 16.380 & 0.000    & 0.000  & ... \\
296.026120   &   -14.685690  & 18.150 & 0.130    & 0.320  & ... & >        & 16.320 & 0.000    & 0.000  & ... \\
...          &   ...         & ...    & ...      & ...    & ... & ...      & ...    & ...      & ...    & ... \\
\midrule 
\end{tabular}
\tablefoot{
d[*]: the differences between the PSF and aperture photometry magnitudes. I\_[*]: ``>'' indicates an upper limit. This table is available in its entirety in CDS. A portion is shown here for guidance regarding its form and content.\\
}
\end{table*}

\begin{table*}
\caption{WISE photometric catalog} 
\scriptsize
\label{wise_cat}
\begin{tabular}{ccccccccccc}
\toprule\toprule
R.A. (J2000) & Decl. (J2000) & [3.4]  & e\_[3.4] & SNR\_[3.4] & ... \\
(deg)        & (deg)         & (mag)  & (mag)    &            & ... \\
\midrule 
295.899753   &   -15.044016  & 15.317 & 0.046    & 23.6       & ... \\
295.899762   &   -14.737916  & 15.846 & 0.065    & 16.8       & ... \\
295.900073   &   -14.581233  & 15.739 & 0.055    & 19.6       & ... \\
...          &   ...         & ...    & ...      & ...        & ... \\
\midrule 
\end{tabular}
\tablefoot{
This table is available in its entirety in CDS. A portion is shown here for guidance regarding its form and content.\\
}
\end{table*}

\section{Additional color-color and color-magnitude diagrams \label{appendix2}}

We present here additional CCDs and CMDs from PS1 versus VHS, PS1 versus IRSF, and PS1 versus HAWK-I, respectively. Each dataset is reduced following the same procedure as described in Section 3.

\begin{figure*}
\center
\includegraphics[scale=0.4]{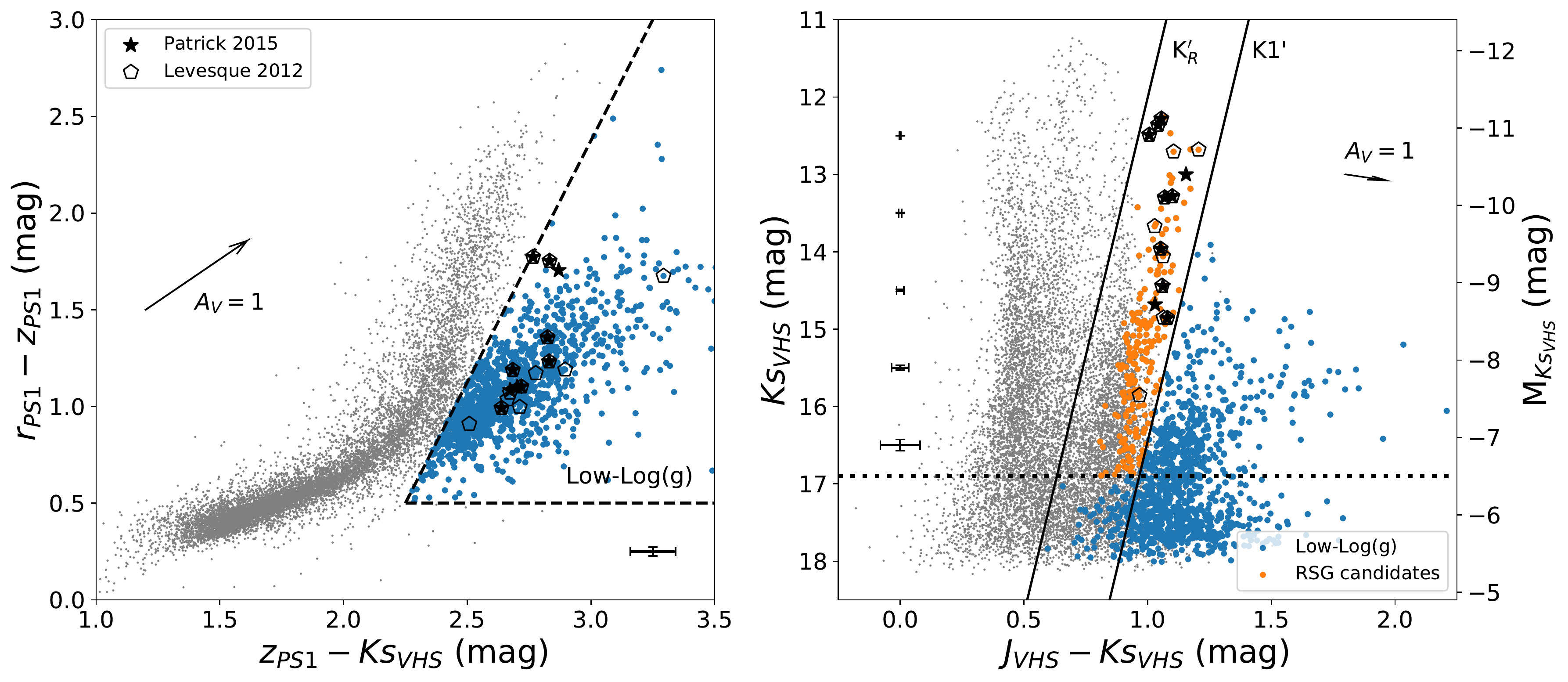}
\caption{Same as Figure~\ref{ngc6822_ccd_ps1_ukirt} and \ref{ngc6822_cmd_ps1_ukirt}, but for combination of PS1 and VHS data. 211 RSG candidates are selected.
\label{ngc6822_ccd_cmd_ps1_vhs}}
\end{figure*}

\begin{figure*}
\center
\includegraphics[scale=0.4]{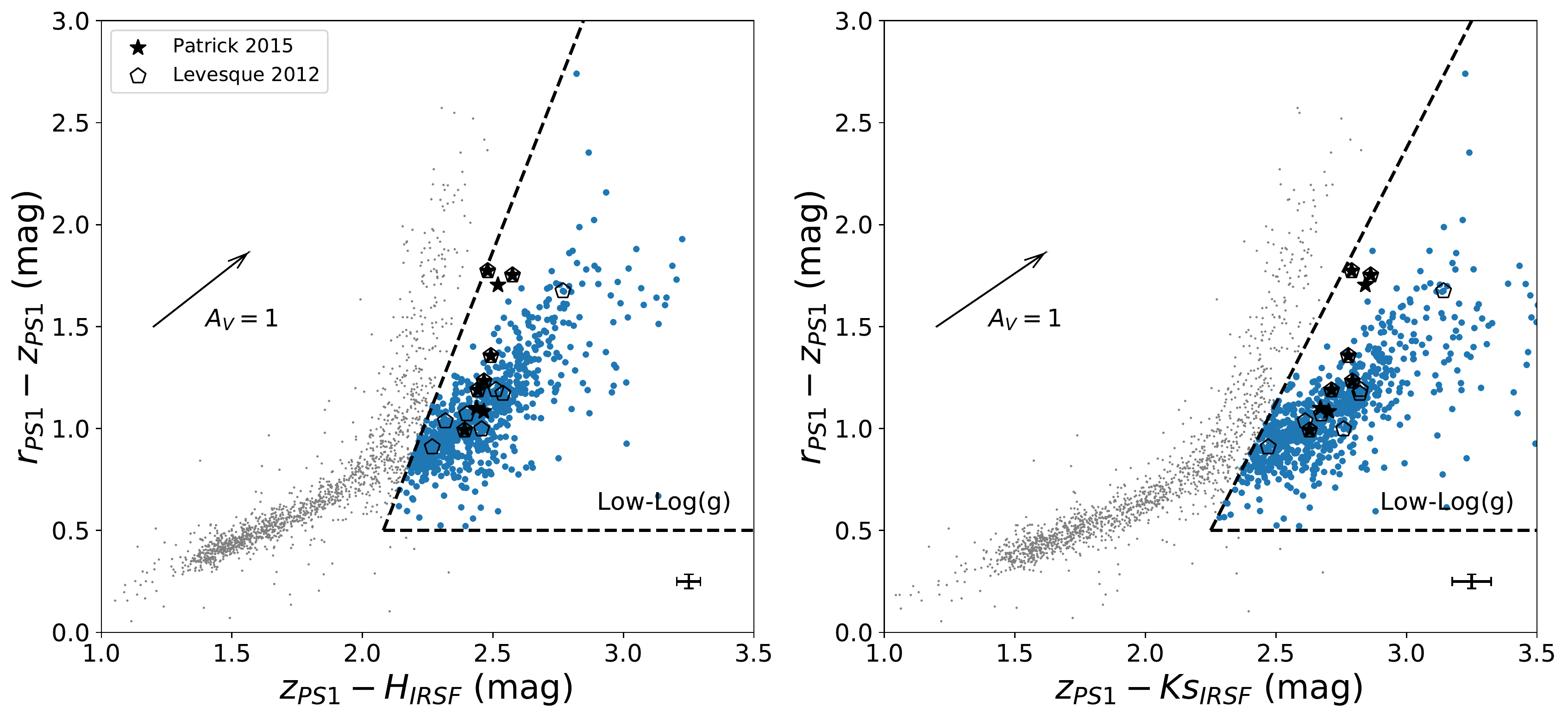}
\includegraphics[scale=0.4]{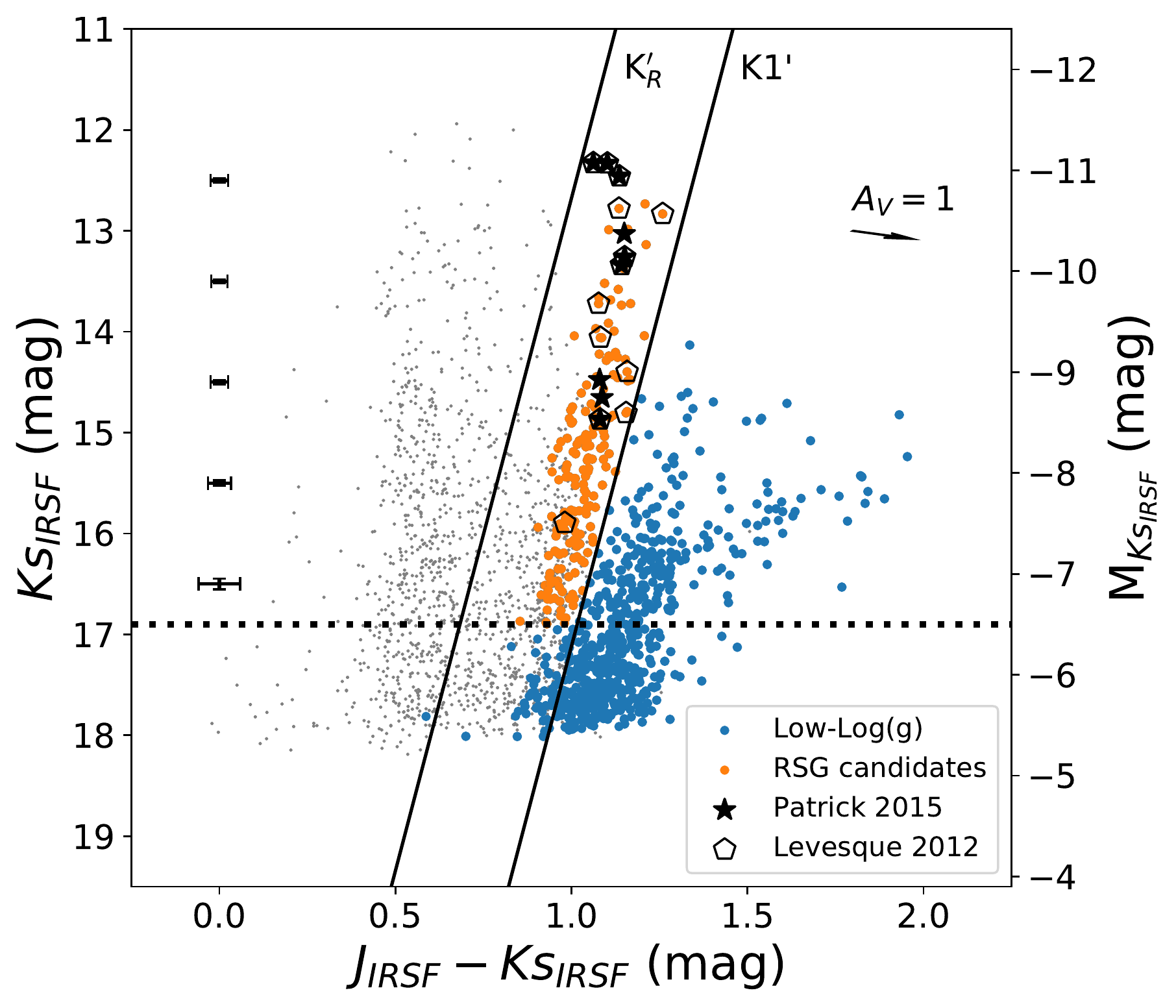}
\caption{Same as Figure~\ref{ngc6822_ccd_ps1_ukirt} and \ref{ngc6822_cmd_ps1_ukirt}, but for combination of PS1 and IRSF data. 183 RSG candidates are selected, where 176 based on rzH, 178 based on rzKs, and 171 in common.
\label{ngc6822_ccd_cmd_ps1_irsf}}
\end{figure*}

\begin{figure*}
\center
\includegraphics[scale=0.4]{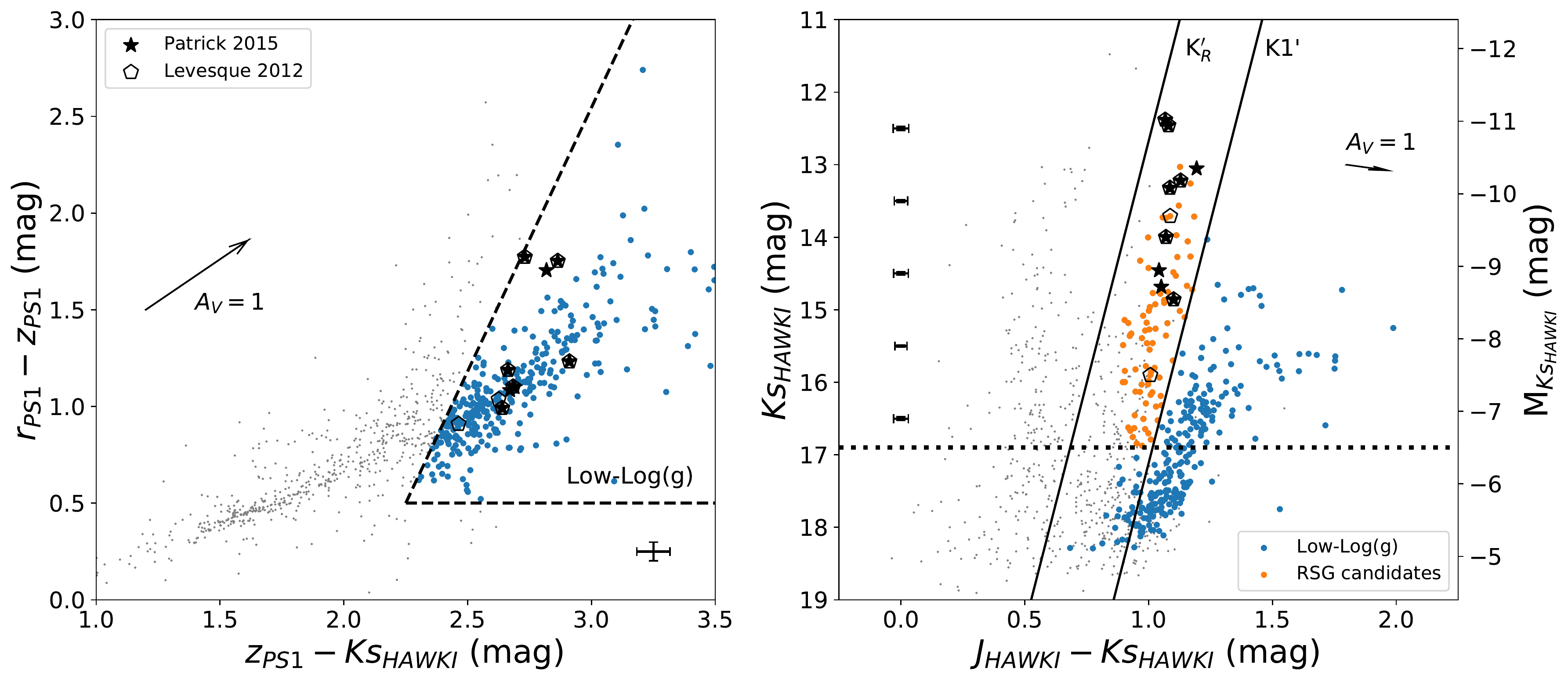}
\caption{Same as Figure~\ref{ngc6822_ccd_ps1_ukirt} and \ref{ngc6822_cmd_ps1_ukirt}, but for combination of PS1 and HWAK-I data. 88 RSG candidates are selected.
\label{ngc6822_ccd_cmd_ps1_hwaki}}
\end{figure*}

\end{appendix}

\clearpage

\end{CJK*}

\end{document}